\documentclass[aps,preprint,showpacs,preprintnumbers,amsmath,amssymb]{revtex4}
\usepackage{amsmath,mathrsfs,amsbsy,color,graphicx,bm,amsthm,amsfonts}
\usepackage{units}
\usepackage{bbm}
\usepackage{times}
\usepackage{dcolumn}
\usepackage{mathrsfs}
\usepackage{amsmath,amssymb,epsfig}
\usepackage{amsmath}
\newcommand{\udots}{\mathinner{\mskip1mu\raise1pt\vbox{\kern7pt\hbox{.}}
\mskip2mu\raise4pt\hbox{.}\mskip2mu\raise7pt\hbox{.}\mskip1mu}}
\begin{document}
\title{Curvature-enhanced multipartite coherence  in the multiverse}
\author{Shu-Min Wu$^1$\footnote{smwu@lnnu.edu.cn}, Chun-Xu Wang$^1$, Rui-Di Wang$^1$, Jin-Xuan Li$^1$, Xiao-Li Huang$^1$\footnote{ huangxiaoli1982@foxmail.com}, Hao-Sheng Zeng $^2$\footnote{hszeng@hunnu.edu.cn (corresponding author)}}
\affiliation{$^1$ Department of Physics, Liaoning Normal University, Dalian 116029, China\\
$^2$ Department of Physics, Hunan Normal University, Changsha 410081, China
}


\begin{abstract}
Here, we study  quantum coherence of N-partite GHZ (Greenberger-Horne-Zeilinger) and W states  in the  multiverse consisting of $N$ causally disconnected de Sitter spaces.
Interestingly, N-partite  coherence  increases monotonically  as the curvature increases,
while  the curvature effect destroys quantum entanglement and discord,
meaning that the curvature effect is beneficial to quantum coherence and harmful to quantum correlations in the multiverse.  We  find that, with the increase of $n$ expanding de Sitter spaces, N-partite coherence of GHZ state increases monotonically for any curvature, while quantum coherence of the W state decreases or increases monotonically depending on the curvature. We  find  a distribution relationship, which indicates that  the  correlated coherence of N-partite W state is equal to the sum of all bipartite correlated coherence in the multiverse. Multipartite coherence exhibits unique properties in the multiverse,  which argues that it  may provide some evidence for the existence of the multiverse.
\end{abstract}

\vspace*{0.5cm}
\maketitle
\section{Introduction}
Quantum coherence, arising from the superposition principle of quantum state, is one of the important features of the quantum world, and is the basis of the fundamental phenomena of quantum interference \cite{L1}.
Like quantum entanglement, quantum coherence is an important quantum resource, which can be applied in quantum information processing, solid state physics, quantum optics, nanoscale thermodynamics, and biological systems \cite{L2,L3,L4,L5,L6,L7,L8,L9,L10,L11,L12}. Although quantum coherence is  of great importance, it did not attract more attention until Baumgratz $et$ $al$. proposed a rigorous resource theory framework for the  quantization of coherence, such as the $l_1$ norm of coherence and the relative entropy of coherence \cite{L13}.  For complex multipartite systems, the $l_1$ norm of coherence
is more directly calculated and is easier to obtain analytical expression than
the relative entropy of coherence. On the other hand, as the quantum information task becomes more and more complex, we need to deal with it with multipartite coherence.

Observer-dependent quantum entanglement can  be discussed
in the background of an expanding universe \cite{L14,L15,L16,L17}.  The
theory of inflationary cosmology and our current observations suggest that
our universe may approach the de Sitter space with a positive cosmological constant in the far past and the far future, which is the unique maximally symmetric
curved spacetime.  Any two mutually separated $R$ and $L$  regions  eventually
are causally disconnected in de Sitter space \cite{L18}, where the universe expands
exponentially.  This is most appropriately
described by spanning open universe coordinates
for two open charts in de Sitter space.  The positive
frequency mode functions of a free massive scalar field correspond to
the Bunch-Davies vacuum (the Euclidean vacuum) that supports on both  $R$ and $L$ regions. Using them, entanglement entropy between two causally
disconnected regions in de Sitter space has been studied in the
Bunch-Davies vacuum and $\alpha$-vacua \cite{L19,L20,L21,L22,L23,L24}. Motivated by
the work, quantum steering, entanglement, and discord were also studied \cite{L25,L26,L27,L28,QL28}.
Since it was shown that quantum entanglement between causally
separated regions (beyond the size of the Hubble horizon) exists in de Sitter space,
there may be the observable effects of quantum
correlations on the cosmic microwave background
(CMB) in our expanding universe.

\begin{figure}
\begin{minipage}[t]{0.5\linewidth}
\centering
\includegraphics[width=3.0in,height=5.2cm]{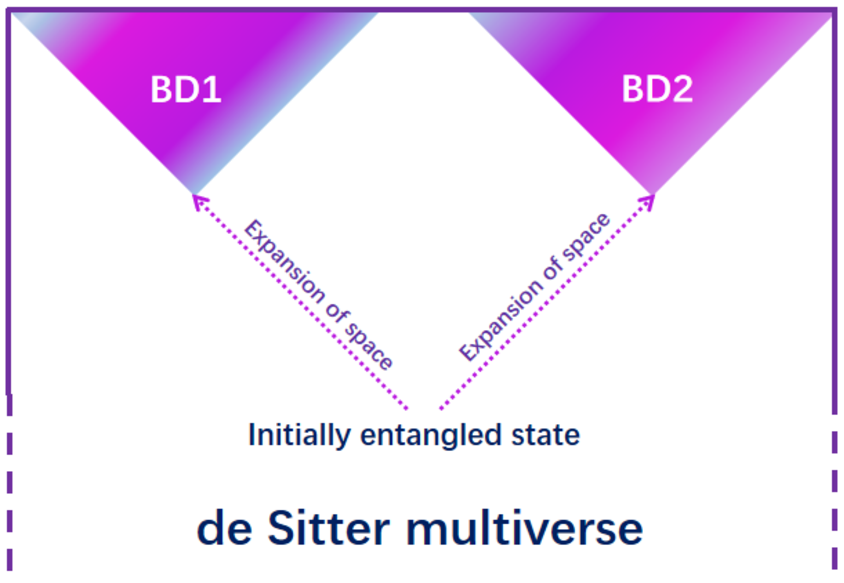}
\end{minipage}%
\caption{Causal diagram of the inflationary multiverse.}
\label{Fig1}
\end{figure}

The vacuum fluctuations in our expanding universe may be entangled with those in another part of the multiverse \cite{L19}. In other words, one considers quantum entanglement between two causally disconnected de Sitter spaces (BD1 and BD2) depicted in Fig.\ref{Fig1} \cite{L26,ZL28}. A quantum  system comprises subsystems BD1 and BD2. Assume that the universe is, say, BD1 and one has no access to BD2. In fact,  quantum entanglement of the reduced density matrix influences the shape
of the spectrum on large scales, which is comparable to or greater than the curvature radius \cite{L20}. This could be the observational signature of the multiverse.
In addition, quantum coherence may be determined by the observers  in the process of bubble nucleation. It is well known that  quantum coherence reflects the nonclassical world better than quantum entanglement, which is considered to be derived from  the nonlocal superposition principle of quantum state. In other words, quantum entanglement is a special kind of quantum coherence (genuine coherence) \cite{L30,L31,L32}.
In general, quantum entanglement and coherence show similar properties in a relativistic setting \cite{L38,L39,ZL1,ZL2,ZL3,ZL4,zht}. It is not clear whether multipartite coherence and quantum entanglement have similar properties in the multiverse.
Therefore, demonstrating the observer's dependence on multipartite coherence in the multiverse is one of the motivations for our work.

Another motivation for our work is  better to understand the multiverse through multipartite coherence. According to the string landscape and inflationary cosmology,  our universe may not be the just one, but part of the multiverse \cite{L33,L34,L35,L36,L37}. In the structure of the multiverse model, there may be many causally disconnected de Sitter bubbles (de Sitter universes). Until recently, the multiverse is merely a philosophical conjecture that has so far been untestable.  However, in the multiverse,  quantum coherence between $N$ causally
separated universes  may generate detectable signatures.
 Some of their quantum states that are far from the
Bunch-Davies vacuum may be entangled with another universes  \cite{L19}.
Then, we  introduce $N$
observers who determine quantum coherence between  $N$ causally disconnected de Sitter
spaces.  We assume $n$ observers  inside de Sitter universes and want to see how the inside $n$ observers detect the signature of quantum coherence with
another $N-n$ de Sitter universes.

In this paper, we  discuss quantum coherence of N-partite GHZ and W
states of massive scalar fields in de Sitter universes.
We assume that $N-n$ observers are in their respective static universes, while $n$ observers are in their expanding universes. Here an observer corresponds to a universe.
We calculate N-partite coherence and obtain  its analytical expression
in de Sitter background. We find that, with
the increase of the curvature, N-partite coherence increases monotonically, while quantum correlation decreases monotonically as the curvature increases in de Sitter universes \cite{L25,L26,L27,L28}. Therefore, we can gain a deeper understanding of the multiverse from the perspective of quantum resources. Although this research may involve quantities that cannot be directly detected in the multiverse, it provides profound insights into the fundamental nature of quantum systems in different spacetime backgrounds.

Interestingly, quantum coherence of N-partite GHZ state in de Sitter universes has
nonlocal coherence and local coherence that can exist in subsystems, while quantum coherence of N-partite GHZ state in Rindler spacetime is genuinely global that cannot exist in any subsystems. We quantify the nonlocal coherence in terms of correlated coherence of the multipartite systems  in the multiverse. N-partite  coherence of W state in de Sitter universes  is not monogamous. However, the correlated coherence of N-partite W state is equal to the sum of the correlated coherence of all the bipartite subsystems  in de Sitter universes. With the increase of the $n$, quantum coherence of W state decreases or increases monotonically depending on curvature, while N-partite coherence of GHZ state increases monotonically for any curvature.
These results show some unique  phenomena of multipartite coherence in de Sitter universes, which provides the possibility for us to find the multiverse.

The paper is organized as follows. In Sec. II, we briefly introduce the quantization of the free massive scalar field in de Sitter space.  In Sec. III, we study quantum coherence of tripartite GHZ and W states in  the multiverse.  In Sec. IV, we extend the relevant research to the N-partite systems.  The last section is devoted to a brief conclusion.

\section{Quantization of scalar field in de Sitter space \label{GSCDGE}}
We consider a free scalar field $\phi$ with the mass $m$ in  the Bunch-Davies vacuum of de Sitter space represented by metric $g_{\mu\nu}$. The action of the field is given by
\begin{eqnarray}\label{w1}
S=\int d^4 x\sqrt{-g}\left[\,-\frac{1}{2}\,g^{\mu\nu}
\partial_\mu\phi\,\partial_\nu \phi
-\frac{m^2}{2}\phi^2\,\right]\,.
\end{eqnarray}
The coordinate systems of the open chart in de Sitter space  can be obtained by analytic continuation from the Euclidean metric and divided into two parts  that we call the $R,L$. The $R$ and $L$ regions, which are covered, respectively, by the coordinates $(t_{L}, r_{L})$ and $(t_{R}, r_{R})$ in de Sitter space,  are causally disconnected, and their metrics are given, respectively, by
\begin{eqnarray}\label{w2}
ds^2_R&=&H^{-2}\left[-dt^2_R+\sinh^2t_R\left(dr^2_R+\sinh^2r_R\,d\Omega^2\right)
\right]\,,\nonumber\\
ds^2_L&=&H^{-2}\left[-dt^2_L+\sinh^2t_L\left(dr^2_L+\sinh^2r_L\,d\Omega^2\right)
\right]\,,
\end{eqnarray}
where  $H^{-1}$ is the Hubble radius and $d\Omega^2$ is the metric on the two-sphere \cite{L18}.  In order to obtain the analytic continuation solutions in the $R$ or $L$ regions, we need to resolve this process in the Euclidean hemisphere.  It is natural to choose  the Euclidean vacuum ( Bunch-Davies vacuum) with de Sitter invariance  as the initial condition. Therefore, it is necessary to find the positive frequency mode functions corresponding to the Euclidean vacuum. By separating the variables, we get
\begin{eqnarray}\label{w3}
\phi=\frac{H}{\sinh t}\chi_{p}(t)Y_{p\ell m}(r,\Omega),
\end{eqnarray}
and the solutions of the Klein-Gordon equations for $\chi_{p}(t)$ and $Y_{p\ell m}(r,\Omega)$ in the $R$ or $L$ regions are found to be
\begin{eqnarray}\label{w4}
\bigg[\frac{\partial^{2}}{\partial t^{2}}+3\coth t \frac{\partial}{\partial t}+\frac{1+p^{2}}{\sinh^{2}t}+\frac{m^{2}}{H^{2}}\bigg]\chi_{p}(t)=0,
\end{eqnarray}
\begin{eqnarray}\label{w5}
\bigg[\frac{\partial^{2}}{\partial r^{2}}+2\coth t \frac{\partial}{\partial r}-\frac{1}{\sinh^{2}t}\mathbf{L^{2}}\bigg]Y_{p\ell m}(r,\Omega)=-(1+p^{2})Y_{p\ell m}(r,\Omega),
\end{eqnarray}
where  $Y_{p\ell m}(r,\Omega)$ are eigenfunctions on the three-dimensional hyperboloid,  and the $\mathbf{L^{2}}$ is the Laplacian operator on the unit two-sphere \cite{L19,QL37}.

The positive frequency mode functions corresponding to
the Euclidean vacuum that is supported both on the $R$ and $L$ regions are given by
\begin{eqnarray}\label{w6}
\chi_{p,\sigma}(t)=\left\{
\begin{array}{l}
\frac{e^{\pi p}-i\sigma e^{-i\pi\nu}}{\Gamma(\nu+ip+\frac{1}{2})}P_{\nu-\frac{1}{2}}^{ip}(\cosh t_R)
-\frac{e^{-\pi p}-i\sigma e^{-i\pi\nu}}{\Gamma(\nu-ip+\frac{1}{2})}P_{\nu-\frac{1}{2}}^{-ip}(\cosh t_R)
\,,\\
\\
\frac{\sigma e^{\pi p}-i\,e^{-i\pi\nu}}{\Gamma(\nu+ip+\frac{1}{2})}P_{\nu-\frac{1}{2}}^{ip}(\cosh t_L)
-\frac{\sigma e^{-\pi p}-i\,e^{-i\pi\nu}}{\Gamma(\nu-ip+\frac{1}{2})}P_{\nu-\frac{1}{2}}^{-ip}(\cosh t_L)
\,,
\label{solutions}
\end{array}
\right.
\end{eqnarray}
where $P^{\pm ip}_{\nu-\frac{1}{2}}$ are the associated   Legendre functions
and  $\sigma=\pm 1$ is used for  distinguishing  the independent solutions in each open region.  Their Klein-Gordon norms are evaluated to give $[\chi_{p,\sigma}(t),\chi_{p,\sigma'}(t)]=N_{p}\delta_{\sigma\sigma'}$ with the normalization factor $N_{p}=\frac{4\sinh\pi p\,\sqrt{\cosh\pi p-\sigma\sin\pi\nu}}{\sqrt{\pi}\,|\Gamma(\nu+ip+\frac{1}{2})|}\,$ \cite{L18}. In the above solutions, $\nu$ is a mass parameter that is given by
\begin{eqnarray}\label{w7}
\nu=\sqrt{\frac{9}{4}-\frac{m^2}{H^2}}\,.
\end{eqnarray}
Note that the  two special values of the mass parameters  $\nu= 1/2$ and $\nu= 3/2$ correspond to the  conformally coupled massless scalar and the minimally coupled massless limit,  respectively. Here,  $p$ is the momentum of the scalar field. The curvature effect in three-dimensional hyperbolic
space starts to appear around $p\sim1$. When the momentum $p$ decreases, the curvature effect becomes stronger. Therefore, we can probe the curvature effect on multipartite coherence by
varying the momentum $p$ \cite{L25,L26,L27,L28,QL28}.

The scalar field can be expanded in terms of the annihilation and creation operators
\begin{eqnarray}\label{w8}
\hat\phi(t,r,\Omega)
=\frac{H}{\sinh t}\int dp \sum_{\sigma,\ell,m}\left[\,a_{\sigma p\ell m}\,\chi_{p,\sigma}(t)
+a_{\sigma p\ell -m}^\dagger\,\chi^*_{p,\sigma}(t)\right]Y_{p\ell m}(r,\Omega)
\,,
\end{eqnarray}
where   $a_{\sigma p\ell m}$ satisfies $a_{\sigma p\ell m}|0\rangle_{\rm BD}=0$ in the Bunch-Davies vacuum. We introduce a Fourier mode field operator
\begin{eqnarray}\label{w9}
\phi_{ p\ell m} (t)\equiv\sum_{\sigma}\left[\,a_{\sigma p\ell m}\,\chi_{p,\sigma}(t)
+a_{\sigma p\ell -m}^\dagger\,\chi^*_{p,\sigma}(t)\right]
\,.
\end{eqnarray}
For simplicity, hereafter we omit the indices $p$, $\ell$, $m$ in the  operators $\phi_{p\ell m}$,
$a_{\sigma p\ell m}$ and $a_{\sigma p\ell -m}^\dag$.
For example, the mode functions and the associated Legendre functions  can be rewritten as
$P_{\nu-1/2}^{ip}(\cosh t_{R,L})\rightarrow P^{R, L}$, $P_{\nu-1/2}^{-ip}(\cosh t_{R,L})\rightarrow P^{R*, L*}$ and $\chi_{p,\sigma}(t)\rightarrow\chi^{\sigma}$.

The two lines of Eq.(\ref{w6}) can be expressed in one line
\begin{eqnarray}
\chi^{\sigma} = \tilde{N}_p^{-1} \sum_{q=R,L} \left[\,
 \alpha_q^\sigma\,P^q + \beta_q^\sigma\,P^{*\,q}
\,\right]\,,
\label{sty2}
\end{eqnarray}
where $ \tilde{N}_p^{-1}=\frac{|\Gamma(1+ip)|}{\sqrt{2p}}$ and
\begin{eqnarray}
&&\alpha_R^\sigma = \frac{e^{\pi p} -i\sigma e^{-i\pi \nu}}{\Gamma (\nu+ip +\frac{1}{2})},\quad
\beta_R^\sigma =-\frac{e^{-\pi p} -i\sigma e^{-i\pi \nu}}{\Gamma (\nu-ip +\frac{1}{2})} \,\,,\\
&&\alpha_L^\sigma =\sigma\,\frac{e^{\pi p} -i\sigma e^{-i\pi \nu}}{\Gamma (\nu+ip +\frac{1}{2})}
\,,\quad
\beta_L^\sigma =-\sigma\,\frac{e^{-\pi p} -i\sigma e^{-i\pi \nu}}{\Gamma (\nu-ip +\frac{1}{2})}   \,\,.
\end{eqnarray}
The complex conjugate of Eq.(\ref{sty2}) reads
\begin{eqnarray}
\chi^{*\,\sigma}=N_p^{-1}\sum_{q=R,L} \left[\,
{\beta^{*}{}_q}^{\!\!\!\sigma}\,P^q + {\alpha^{*}{}_q}^{\!\!\!\sigma}\,P^{*\,q}
\,\right]\,.
\end{eqnarray}
Then, Eq.(\ref{w6}) and its conjugate can be put into the simple matrix form \cite{L22,L26}
\begin{eqnarray}
\chi^I=N_p^{-1}\,M^I{}_J\,P^J\,,
\end{eqnarray}
where the capital indices $(I,J)$ run from 1 to 4 and
\begin{eqnarray}
\chi^I=\left(\,\chi^\sigma\,,\chi^{*\,\sigma}\,\right)\,,\quad
M^I{}_J=\left(
\begin{array}{ll}
\alpha^\sigma_q & \beta^\sigma_q \vspace{3mm}\\
{\beta^{*}{}_q}^{\!\!\!\sigma} & {\alpha^{*}{}_q}^{\!\!\!\sigma} \\
\end{array}\right)\,,\quad
P^J=\left(\,P^R\,,P^L\,,P^{*\,R}\,, P^{*\,L}\,\right)\,.
\end{eqnarray}
Now we introduce the new  annihilation and creation operators
($b_q,b_q^\dag$) that satisfy $b_q|0\rangle_{q}=0$ in different regions.
Because  the Fourier mode field operator is the same under the change of mode functions, we can relate the annihilation and creation operators ($b_q,b_q^\dag$) and $(a_\sigma,a_\sigma^\dag)$  in different reference  \cite{L22,L26}. Therefore, we have
\begin{eqnarray}
\phi(t)=a_I\,\chi^I=N_p^{-1}a_I\,M^I{}_J\,P^J\,=N_p^{-1}b_J\,P^J\,,\quad
a_I=\left(\,a_\sigma\,,\,a_\sigma^\dagger\,\right)\,,\quad b_J=\left(\,b_R\,,\,b_L\,,\,b_R^\dagger\,,\, b_L^\dagger\,\right)\,.
\label{phi2}
\end{eqnarray}
From Eq.(\ref{phi2}), we obtain the relation as
\begin{eqnarray}
a_J=b_I\left(M^{-1}\right)^I{}_J\,,\qquad
\left(M^{-1}\right)^I{}_J=\left(
\begin{array}{ll}
\xi_{q\sigma} & \delta_{q\sigma} \vspace{3mm}\\
\delta_{q\sigma}^* & \xi_{q\sigma}^* \\
\end{array}\right)\,,\qquad
\left\{
\begin{array}{l}
\xi=
\left(\alpha-\beta\,\alpha^{*\,-1}\beta^*\right)^{-1}\,,\vspace{3mm}\\
\delta=-\alpha^{-1}\beta\,\xi^*\,.
\end{array}
\right.
\label{xidelta1}
\end{eqnarray}

Using the Bogoliubov transformation between the operators, the Bunch-Davies vacuum and single particle excitation states can be constructed from the
states in  $R$ and $L$ regions \cite{L28}, which can be expressed as
\begin{eqnarray}\label{w12}
|0\rangle_{\rm BD} = \sqrt{1-|\gamma_{p}|^{2}}\sum_{n=0}^{\infty}\gamma_{p}^{n}|n\rangle_{L}|n\rangle_{R}\,,
\label{bogoliubov3}
\end{eqnarray}
\begin{eqnarray}\label{w13}
|1\rangle_{\rm BD}=\frac{1-|\gamma_{p}|^{2}}{\sqrt{2}}\sum_{n=0}^{\infty}\gamma^{n}_{p}\sqrt{n+1}\big[|(n+1)\rangle_{\rm L}|n\rangle_{\rm R}+|n\rangle_{\rm L}|(n+1)\rangle_{\rm R}\big],
\label{bogoliubov3}
\end{eqnarray}
where $|n\rangle_{R}$ and $|n\rangle_{L}$ correspond to the two modes of the $R$ and $L$ de Sitter open charts, respectively, and the parameter $\gamma_p$ reads
\begin{eqnarray}\label{w14}
\gamma_p = i\frac{\sqrt{2}}{\sqrt{\cosh 2\pi p + \cos 2\pi \nu}
 + \sqrt{\cosh 2\pi p + \cos 2\pi \nu +2 }}\,.
\label{gammap2}
\end{eqnarray}

Now, we elaborate on the assertion that $p$ can be regarded as the curvature parameter of the de Sitter space. Employing Eq.(\ref{w12}), the reduced density matrix from the Bunch-Davies basis to the basis of the open chart in the $L$ region can be expressed as
\begin{eqnarray}\label{pom1}
\rho_{\rm L}=\rm{Tr_R(|0\rangle_{BD}\langle0|)}=(1-|\gamma_{p}|^{2})\sum_{n=0}^{\infty}|\gamma_{p}|^{2n}| n\rangle_{\rm L}\langle  n|.
\end{eqnarray}
Similarly, we can also obtain the reduced density matrix  from the Minkowski basis to the
Rindler basis that  is found to be
\begin{eqnarray}\label{pom2}
\rho_{\rm I}=(1-e^{-2\pi \frac{\omega}{a}})\sum_{n=0}^{\infty}e^{-2\pi n\frac{\omega}{a}}| n\rangle_{\rm I}\langle  n|,
\end{eqnarray}
where $a$ is the acceleration \cite{ZL1,ZL3}.  From Eq.(\ref{pom1}) and Eq.(\ref{pom2}), we can obtain $a=-\pi\omega/\ln|\gamma_{p}|$. For the given values of $\nu$ and $\omega$, it is evident that the temperature $T_U=\frac{a}{2\pi}$ is a monotonically decreasing function of $p$. That is to say, a decrease in $p$ will lead to an increase in curvature effect of the de Sitter space. This assertion has been utilized previously \cite{L25,L26,L27,L28,QL28}.

\section{Tripartite  coherence of scalar fields in the multiverse \label{GSCDGE}}
Utilizing this Bogoliubov transformation given by Eqs.(\ref{w12}) and (\ref{w13}), we discover that the initial state of the Bunch-Davies mode observed by an observer in the global chart corresponds to a two-mode squeezed state in the open charts.
These two modes correspond to the fields observed in the $R$ and $L$ charts. If we exclusively examine one of the open charts, let's say $L$, we cannot access to the modes in the causally disconnected $R$ region and must consequently trace over the inaccessible region.
This situation is analogous to the relationship between an observer in a Minkowski chart and another in one of the two Rindler charts. In this sense, the global chart and Minkowski chart encompass the entire spacetime geometry, while the open charts and Rindler charts cover only a portion of the spacetime. From the above analysis, we find that the time evolution does not play a direct role in the calculations presented in the paper. The focus is primarily on the initial quantum states and the subsequent tracing out of parts of the density matrix.

In the structure of the multiverse model, there may be many causally disconnected de Sitter bubbles (de Sitter universes), and the inside of a nucleated bubble looks like an open universe. Along this line, we initially consider two typical tripartite GHZ and W states shared by Alice, Bob, and Charlie who determine quantum coherence between three causally disconnected de Sitter spaces. The technology for preparing GHZ and W states in experiments has become very mature \cite{plm1,plm2}.  Quantum coherence can be observed experimentally  in flat spacetime \cite{FGH1,FGH2}.
As is well known, the search for multiverse observations remains an open question, and most of the research on multiverse is theoretical \cite{L26,ZL28}. Therefore, we believe that understanding the behavior of quantum coherence of GHZ and W states can provide guidance for simulating multiverse using quantum systems \cite{ZQQL1,ZQQL2,ZQQL3}.
 By removing a single particle from the W state, the ensuing bipartite state remains entangled. Thus, the W state demonstrates a remarkable persistence of quantum entanglement in the face of particle loss. Unlike W state, quantum entanglement and coherence of GHZ state only exist in three particles.
Then, we set  Bob and Charlie, respectively, to stay in  the $L$ regions of  two expanded de Sitter universes, and
Alice is in the global chart of the other de Sitter space without expansion. In experiments,
it is temporarily difficult to realistically place an entangled particle in every universe, but this is just a setting in our theoretical model. If we probe only an open chart, such as $L$, the observer cannot access the mode in the causally disconnected $R$ region,  and the inaccessible $R$  region must be  traced over. In other words, in composite quantum systems, we only focus on the modes under consideration, so we need to trace the remaining modes.
Then a pure state of the observers is going to be a mixed state.  Therefore,
thermal noise introduced by the expanding universe destroys quantum correlations \cite{L25,L26,L27,L28}. In the following, we will explore the properties of tripartite coherence in the multiverse.

\subsection{Tripartite GHZ state }
We assume that three observers,  Alice, Bob, and Charlie share a tripartite GHZ state of  the free massive scalar field defined as
\begin{eqnarray}\label{w15}
|GHZ\rangle_{ABC}=\frac{1}{\sqrt{2}}[|0\rangle_{A, \rm BD_{1}}|0\rangle_{B, \rm BD_{2}}|0\rangle_{C, \rm BD_{3}}+|1\rangle_{A, \rm BD_{1}}|1\rangle_{B, \rm BD_{2}}|1\rangle_{C, \rm BD_{3}}].
\end{eqnarray}
Here, Alice, Bob, and Charlie are in  the  three causally disconnected de Sitter spaces $(\rm BD_{1})$, $(\rm BD_{2})$ and $(\rm BD_{3})$, respectively. Then,
we consider that Bob and Charlie are in the $L$ regions of expanded de Sitter universes and Charlie is in a global chart of the other de Sitter space. For convenience, we omit the subscript $\rm{BD}$. Using Eqs.(\ref{w12}) and (\ref{w13}), we can rewrite Eq.(\ref{w15}) as
\begin{eqnarray}\label{w16}
\begin{aligned}
|GHZ\rangle_{A B \bar{B} C \bar{C}}=&\frac{1-|\gamma_{p}|^{2}}{\sqrt{2}}\sum_{n,m=0}^{\infty}\gamma^{n}_{p}\gamma^{m}_{p}\bigg[|0\rangle_A |n\rangle_{B}| n\rangle_{\bar{B}}|m\rangle_{C}|m\rangle_{\bar{C}}\\
+&\frac{1-|\gamma_{p}|^{2}}{2}\sqrt{n+1}\sqrt{m+1}|1\rangle_A(|n+1\rangle_{B}|n\rangle_{\bar{B}}+|n\rangle_{B}|n+1\rangle_{\bar{B}})\\
\otimes&(|m+1\rangle_{C}|m\rangle_{\bar{C}}+|m\rangle_{C}|m+1\rangle_{\bar{C}})\bigg],
\end{aligned}
\end{eqnarray}
where the modes $\bar{B}$ and $\bar{C}$ are in the  $R$ regions.  Since the $R$ and $L$ regions are causally disconnected,  we need to trace over the modes $\bar{B}$ and $\bar{C}$ in the $R$ regions and obtain the reduced density matrix  as
\begin{eqnarray}\label{w17}
\rho_{ABC}=\frac{(1-|\gamma_{p}|^{2})^{2}}{2}\sum_{n,m=0}^{\infty}\gamma_{p}^{2n}\gamma_{p}^{2m}\rho_{n,m},
\end{eqnarray}
where
\begin{equation}\label{w18}
\begin{aligned}
\rho_{n,m}=&|0\rangle_{A}\langle0||n\rangle_{B}\langle n||m\rangle_{C}\langle m|+\frac{1-|\gamma_{p}|^{2}}{2}\sqrt{n+1}\sqrt{m+1}|0\rangle_{A}\langle1|\\ \otimes&(|n\rangle_{B}\langle n+1|+\gamma_{p}|n+1\rangle_{B}\langle n|)(|m\rangle_{C}\langle m+1|+\gamma_{p}|m+1\rangle_{C}\langle m|)\\+&\frac{1-|\gamma_{p}|^{2}}{2}\sqrt{n+1}\sqrt{m+1}|1\rangle_{A}\langle0|(|n+1\rangle_{B}\langle n|+\gamma_{p}^{*}|n\rangle_{B}\langle n+1|)\\ \otimes&(|m+1\rangle_{C}\langle m|+\gamma_{p}^{*}|m\rangle_{C}\langle m+1|)+\frac{(1-|\gamma_{p}|^{2})^{2}}{4}|1\rangle_{A}\langle1|\\ \otimes&[(n+1)(|n+1\rangle_{B}\langle n+1|+|n\rangle_{B}\langle n|)+\sqrt{n+2}\sqrt{n+1}\\ \times&(\gamma_{p}|n+2\rangle_{B}\langle n|+\gamma_{p}^{*}|n\rangle_{B}\langle n+2|)] [(m+1)(|m+1\rangle_{C}\langle m+1|+|m\rangle_{C}\langle m|)\\+&\sqrt{m+2}\sqrt{m+1}(\gamma_{p}|m+2\rangle_{C}\langle m|+\gamma_{p}^{*}|m\rangle_{C}\langle m+2|)].
\end{aligned}
\end{equation}

Next, we will explore the properties of quantum coherence of the GHZ state in the multiverse. Here, we use the $l_{1}$ norm of coherence introduced by Baumgratz $et$ $al.$ to quantify quantum coherence \cite{L13}, which is defined as the sum of the absolute value of all the off-diagonal elements of a system density matrix,
\begin{eqnarray}\label{w19}
C(\rho)=\sum_{i\neq j}|\rho_{i,j}|.
\end{eqnarray}
It is essential to emphasize that quantum coherence is contingent upon the selection of a reference basis. Representing the same quantum state with different reference bases can result in different values of quantum coherence. In practical, the selection of the reference basis may be governed by the physics inherent in the problem under consideration.
For instance, one might concentrate on the energy eigenbasis when exploring coherence in the context of transport phenomena and thermodynamics. In the quantum depiction of Young's two-slit interference, the path basis is advantageous. In this paper, we employ the particle number representation to investigate the dynamics of multipartite coherence for scalar fields in the multiverse. In quantum optics, the coherent superposition of number states with different numbers of photons is crucial, playing significant roles in various optical interference settings \cite{L123}.
As is widely recognized, coherent and squeezed states of optical fields stand out as typical examples of such coherent superposition. In the case of two-mode optical fields, coherent superposition in the photon-number bases can lead to the emergence of entanglement between the photons of the two modes, such as in the case of two-mode squeezed states. This represents an essential resource in various applications. The coherence resulting from the superposition of photon-number states typically leads to a nonuniform distribution of optical intensity concerning position or time. This phenomenon gives rise to interference fringes, which can be detected through appropriate settings. The maturity of technologies in quantum optics provides the foundation for us to investigate counterparts for scalar fields in the multiverse.

Employing Eqs.(\ref{w17}) and (\ref{w19}), quantum coherence of the GHZ state becomes
\begin{equation}\label{w20}
\begin{aligned}
C(\rho_{ABC})=&\frac{1}{2}\big\{(1-|\gamma_{p}|^{2})^{3}\big[\sum_{n=0}^{\infty}\gamma_{p}^{2n}\sqrt{n+1}(1+|\gamma_{p}|)\big]^{2}\\+&
[(1-|\gamma_{p}|^{2})^{2}\sum_{n=0}^{\infty}\gamma_{p}^{2n+1}\sqrt{n+2}\sqrt{n+1}]^{2}\\
+&2(1-|\gamma_{p}|^{2})^{2}\sum_{n=0}^{\infty}\gamma_{p}^{2n+1}\sqrt{n+2}\sqrt{n+1} \}.
\end{aligned}
\end{equation}
In the above calculations, we have used the relations
\begin{equation}\label{w21}
(1-|\gamma_{p}|^{2})\sum_{n=0}^{\infty}|\gamma_{p}|^{2n}=1,
\end{equation}
and
\begin{equation}\label{w22}
(1-|\gamma_{p}|^{2})^{2}\sum_{n=0}^{\infty}|\gamma_{p}|^{2n}(n+1)=1.
\end{equation}
From  Eq.(\ref{w20}), we can see that quantum coherence  depends on the curvature parameter $p$ and the mass parameter $\nu$, which means that the curvature effect affects quantum coherence in de Sitter spaces.

\begin{figure}
\begin{minipage}[t]{0.5\linewidth}
\centering
\includegraphics[width=3.0in,height=5.2cm]{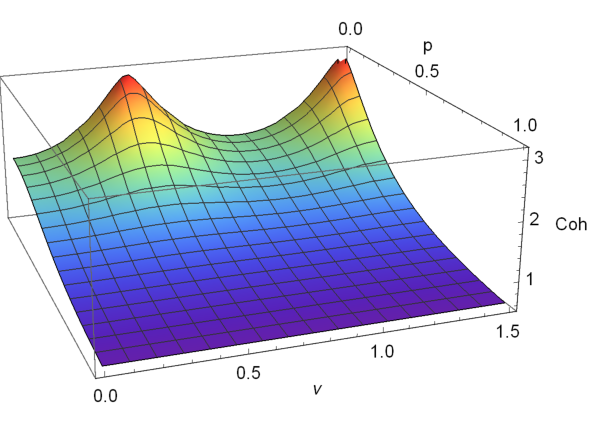}
\end{minipage}%
\caption{Quantum coherence $C(\rho_{ABC})$ of the GHZ state as  functions of the mass parameter $\nu$ and the curvature parameter $p$.}
\label{Fig2}
\end{figure}

In Fig.\ref{Fig2}, we  plot   quantum coherence $C(\rho_{ABC})$ as  functions of the mass parameter $\nu$ and the curvature parameter $p$. From Fig.\ref{Fig2}, we can see that  quantum coherence of the GHZ state decreases monotonically with the increase of the curvature parameter $p$. In other words, quantum coherence of the GHZ state increases with the increase of the curvature. This means that the curvature effect can improve quantum coherence  in the multiverse. Conversely, quantum entanglement  and discord decrease with the increase of the curvature in de Sitter spaces \cite{L25,L26,L27,L28}.
In addition, quantum coherence of the GHZ state is more sensitive to the curvature effect for $\nu\rightarrow1/2$ (conformally coupled massless) and $\nu\rightarrow3/2$ ( minimally coupled massless limit).

Tracing over the modes $A$, $B$ or $C$ from $\rho_{ABC}$, respectively, we obtain
\begin{equation}\label{w23}
\begin{aligned}
\rho_{BC}=&\frac{1}{2}\big\{ (1-|\gamma_{p}|^{2})^{2}\sum_{n=0}^{\infty}\gamma_{p}^{2n}|n\rangle_{B}\langle n|\otimes(1-|\gamma_{p}|^{2})^{2}\sum_{m=0}^{\infty}\gamma_{p}^{2m}|m\rangle_{C}\langle m|\\+&\sum_{n=0}^{\infty}\big[\frac{(1-|\gamma_{p}|^{2})^{2}}{2}\gamma_{p}^{2n}(n+1)(|n\rangle_{B}\langle n|+|n+1\rangle_{B}\langle  n+1|)\\+&\frac{(1-|\gamma_{p}|^{2})^{2}}{2}\gamma_{p}^{2n}\sqrt{n+1}\sqrt{n+2}(\gamma_{p}|n+2\rangle_{B}\langle n|+\gamma_{p}^{*}|n\rangle_{B}\langle n+2|)\big]\\ \otimes&\sum_{m=0}^{\infty}\big[\frac{(1-|\gamma_{p}|^{2})^{2}}{2}\gamma_{p}^{2m}(m+1)(|m\rangle_{C}\langle m|+|m+1\rangle_{C}\langle  m+1|)\\+&\frac{(1-|\gamma_{p}|^{2})^{2}}{2}\gamma_{p}^{2m}\sqrt{m+1}\sqrt{m+2}(\gamma_{p}|m+2\rangle_{C}\langle m|+\gamma_{p}^{*}|m\rangle_{C}\langle m+2|)\big]\big\},
\end{aligned}
\end{equation}
\begin{equation}\label{w24}
\begin{aligned}
\rho_{AC}=&
\frac{1}{2}\big\{|0\rangle_{A}\langle0|\otimes(1-|\gamma_{p}|^{2})^{2}\sum_{m=0}^{\infty}\gamma_{p}^{2m}|m\rangle_{C}\langle m|\\+&|1\rangle_{A}\langle1| \otimes \sum_{m=0}^{\infty}\big[(\frac{(1-|\gamma_{p}|^{2})^{2}}{2}\gamma_{p}^{2m}(m+1)|m\rangle_{C}\langle m|+|m+1\rangle_{C}\langle  m+1|)\\+& \frac{(1-|\gamma_{p}|^{2})^{2}}{2}\gamma_{p}^{2m}\sqrt{m+1}\sqrt{m+2}(\gamma_{p}|m+2\rangle_{C}\langle m|+\gamma_{p}^{*}|m\rangle_{C}\langle m+2|)\big]\big\},
\end{aligned}
\end{equation}
\begin{equation}\label{w25}
\begin{aligned}
\rho_{AB}=&
\frac{1}{2}\big\{|0\rangle_{A}\langle0|\otimes(1-|\gamma_{p}|^{2})^{2}\sum_{m=0}^{\infty}\gamma_{p}^{2m}|m\rangle_{B}\langle m|\\+&|1\rangle_{A}\langle1| \otimes \sum_{m=0}^{\infty}\big[(\frac{(1-|\gamma_{p}|^{2})^{2}}{2}\gamma_{p}^{2m}(m+1)|m\rangle_{B}\langle m|+|m+1\rangle_{B}\langle  m+1|)\\+& \frac{(1-|\gamma_{p}|^{2})^{2}}{2}\gamma_{p}^{2m}\sqrt{m+1}\sqrt{m+2}(\gamma_{p}|m+2\rangle_{B}\langle m|+\gamma_{p}^{*}|m\rangle_{B}\langle m+2|)\big]\big\}. \end{aligned}
\end{equation}
Similarly, we get the density matrixs of the subsystem $A$, $B$ and $C$
\begin{equation}\label{w26}
\rho_{A}=\frac{1}{2}(|0\rangle_{A}\langle0|+|1\rangle_{A}\langle1|),
\end{equation}
\begin{equation}\label{w27}
\begin{aligned}
\rho_{B}=&\frac{1}{2}\big\{|0\rangle_{B}\langle0|+\sum_{m=0}^{\infty}\big[(\frac{(1-|\gamma_{p}|^{2})^{2}}{2}\gamma_{p}^{2m}(m+1)|m\rangle_{B}\langle m|+|m+1\rangle_{B}\langle  m+1|)\\+& \frac{(1-|\gamma_{p}|^{2})^{2}}{2}\gamma_{p}^{2m}\sqrt{m+1}\sqrt{m+2}(\gamma_{p}|m+2\rangle_{B}\langle m|+\gamma_{p}^{*}|m\rangle_{B}\langle m+2|)\big]\big\},
\end{aligned}
\end{equation}
\begin{equation}\label{w28}
\begin{aligned}
\rho_{C}=&\frac{1}{2}\big\{|0\rangle_{C}\langle0|+\sum_{m=0}^{\infty}\big[(\frac{(1-|\gamma_{p}|^{2})^{2}}{2}\gamma_{p}^{2m}(m+1)|m\rangle_{C}\langle m|+|m+1\rangle_{C}\langle  m+1|)\\+& \frac{(1-|\gamma_{p}|^{2})^{2}}{2}\gamma_{p}^{2m}\sqrt{m+1}\sqrt{m+2}(\gamma_{p}|m+2\rangle_{C}\langle m|+\gamma_{p}^{*}|m\rangle_{C}\langle m+2|)\big]\big\}.
\end{aligned}
\end{equation}
Using Eq.(\ref{w19}), we obtain  bipartite coherence as
\begin{equation}\label{w29}
C(\rho_{AB})=C(\rho_{AC})=\frac{(1-|\gamma_{p}|^{2})^{2}}{2}\sum_{n=0}^{\infty}\sqrt{n+2}\sqrt{n+1}|\gamma_{p}|^{2n+1},
\end{equation}
\begin{equation}\label{w30}
\begin{aligned}
C(\rho_{BC})=&\frac{1}{2}\big \{[(1-|\gamma_{p}|^{2})^{2}\sum_{n=0}^{\infty}\sqrt{n+2}\sqrt{n+1}
|\gamma_{p}|^{2n+1}]^{2}\\+&2(1-|\gamma_{p}|^{2})^{2}\sum_{n=0}^{\infty}\sqrt{n+2}\sqrt{n+1}
|\gamma_{p}|^{2n+1}\big\}.
\end{aligned}
\end{equation}
From Eqs.(\ref{w26})-(\ref{w28}),  we can find $C(\rho_{A})$ = 0 and $C(\rho_{AB})
= C(\rho_{AC})=C(\rho_{B})=C(\rho_{C})$.  This means that  quantum coherence $C(\rho_{A})$ in the global  chart of de Sitter space is always zero, while quantum coherence $C(\rho_{B})$ and $C(\rho_{C})$ in the $L$ region of de Sitter space can be generated by the curvature effect.

\begin{figure}
\begin{minipage}[t]{0.5\linewidth}
\centering
\includegraphics[width=3in]{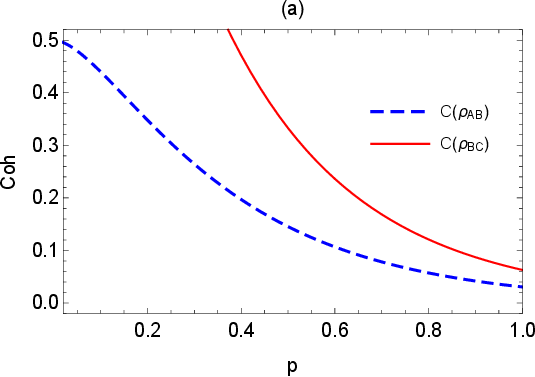}
\label{fig:side:a}
\end{minipage}%
\begin{minipage}[t]{0.5\linewidth}
\centering
\includegraphics[width=3in]{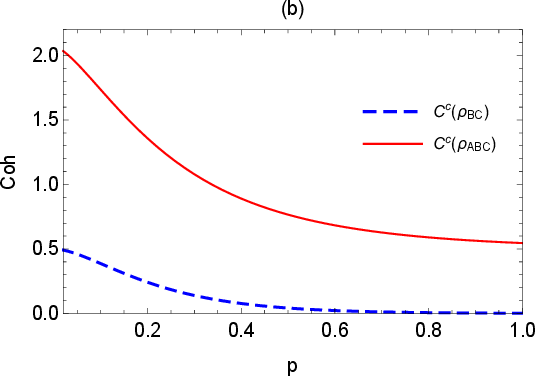}
\label{fig:side:a}
\end{minipage}%

\caption{Bipartite coherence  $C(\rho_{AB})$, $C(\rho_{BC})$, correlated coherence $C^{c}(\rho_{BC})$ and $C^{c}(\rho_{ABC})$  of the GHZ state as a function of the curvature parameter $p$ for a fixed $\nu=3/2$.
 }
\label{Fig3}
\end{figure}

By calculation,  we obtain an inequality  $C(\rho_{BC})> C(\rho_{B}) + C(\rho_{C})$, meaning that the curvature effect can generate nonlocal coherence between the modes $B$ and $C$ in de Sitter spaces. One can define the correlated coherence of a multipartite quantum system described by the density operator $\rho_{A_{1}\cdot\cdot\cdot A_{n}}$ \cite{L30,L31,L32}, which can be expressed as
\begin{equation}\label{w31}
\begin{aligned}
C^{c}(\rho_{A_{1}\cdot\cdot\cdot A_{n}}):=C(\rho_{A_{1}\cdot\cdot\cdot A_{n}}) -\sum_{i}^{n}C(\rho_{A_{i}}).
\end{aligned}
\end{equation}
We obtain the correlated coherence as
\begin{equation}
C^{c}(\rho_{BC})=\frac{1}{2}[(1-|\gamma_{p}|^{2})^{2}\sum_{n=0}^{\infty}\sqrt{n+2}\sqrt{n+1}
|\gamma_{p}|^{2n+1}]^{2},
\end{equation}
\begin{equation}\label{w32}
\begin{aligned}
C^{c}(\rho_{ABC})=&\frac{1}{2}\big\{(1-|\gamma_{p}|^{2})^{3}\big[\sum_{n=0}^{\infty}\gamma_{p}^{2n}\sqrt{n+1}(1+|\gamma_{p}|
)\big]^{2}\\&
+(1-|\gamma_{p}|^{2})^{4}\big[\sum_{n=0}^{\infty}\gamma_{p}^{2n+1}\sqrt{n+1}\sqrt{n+2}\big]^{2}
\big\}.
\end{aligned}
\end{equation}
From Fig.\ref{Fig3}, we find that the bipartite (correlated) coherence decreases with the increase of the curvature parameter $p$, meaning that the curvature effect can generate the bipartite (correlated) coherence.

\subsection{Tripartite W state }
\begin{figure}
\begin{minipage}[t]{0.5\linewidth}
\centering
\includegraphics[width=3.0in,height=5.2cm]{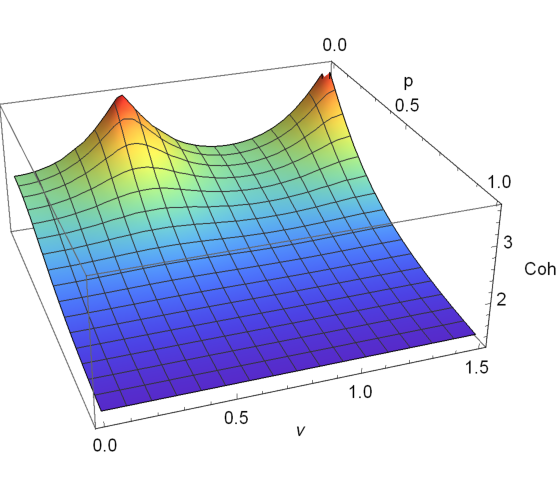}
\end{minipage}%
\caption{Quantum coherence $C(\rho_{ABC})$ of the W state as  functions of the mass parameter $\nu$ and the curvature parameter $p$.}
\label{Fig4}
\end{figure}

In this section, we assume that initially Alice, Bob, and
Charlie share the W state
\begin{eqnarray}\label{w33}
|W\rangle&=&\frac{1}{\sqrt{3}}[|0\rangle_{A, \rm BD_{1}}|0\rangle_{B, \rm BD_{2}}|1\rangle_{C, \rm BD_{3}}
+|0\rangle_{A, \rm BD_{1}}|1\rangle_{B, \rm BD_{2}}|0\rangle_{C, \rm BD_{3}} \\ \nonumber
&+&|1\rangle_{A, \rm BD_{1}}|0\rangle_{B, \rm BD_{2}}|0\rangle_{C, \rm BD_{3}}].
\end{eqnarray}
Following the treatment for the GHZ state, the W state becomes
\begin{eqnarray}\label{w34}
\begin{aligned}
|W\rangle_{AB\bar{B}C\bar{C}}=&\frac{(1-|\gamma_{p}|^{2})^{\frac{3}{2}}}{\sqrt{6}}\sum_{n,m=0}^{\infty}\gamma_{p}^{n}\gamma_{p}^{m}\big[\sqrt{m+1}|0\rangle_{A} |n\rangle_{B} |n\rangle_{\bar{B}}(| m\rangle_{C}|m+1\rangle_{\bar{C}}\\+&|(m+1)\rangle_{C}|m\rangle_{\bar{C}} )+\sqrt{n+1}|0\rangle_{A}(|n\rangle_{B}|n+1\rangle_{\bar{B}}+|n+1\rangle_{B}+|n\rangle_{\bar{B}})\\ \otimes&|m\rangle_{C}|m\rangle_{\bar{C}}+\sqrt{2}(1-|\gamma_{p}|^{2})^{-\frac{1}{2}}|1\rangle_{A} |n\rangle_{B}|n\rangle_{\bar{B}} |m\rangle_{C}|m\rangle_{\bar{C}}\big].
\end{aligned}
\end{eqnarray}
The reduced density matrix after tracing over the modes $\bar{B}$ and $\bar{C}$ can be expressed as
\begin{eqnarray}\label{w35}
\rho_{ABC}=\frac{1}{3}\sum_{n,m=0}^{\infty}\gamma_{p}^{2n}\gamma_{p}^{2m}\rho_{n,m}
\end{eqnarray}
\begin{equation}\label{w36}
\begin{aligned}
\rho_{n,m}=&|0\rangle_{A}\langle0|\otimes(1-|\gamma_{p}|^{2})|n\rangle_{B}\langle n| \otimes \frac{(1-|\gamma_{p}|^{2})^{2}}{2}[(m+1)(|m\rangle_{C}\langle m|+|m+1\rangle_{C}\langle m+1|)
\\+&\sqrt{m+2}\sqrt{m+1}(\gamma_{p}^{*}|m\rangle_{C}\langle m+2|+\gamma_{p}|m+2\rangle_{C}\langle m|)]
\\+&|0\rangle_{A}\langle0|\otimes\frac{(1-|\gamma_{p}|^{2})^{2}}{2}[(n+1)(|n\rangle_{B}\langle n|+|n+1\rangle_{B}\langle n+1|)\\+&\sqrt{n+1}\sqrt{n+2} (\gamma_{p}^{*}|n\rangle_{B}\langle n+2|+\gamma_{p}|n+2\rangle_{B}\langle n|)](1-|\gamma_{p}|^{2})\otimes|m\rangle_{C}\langle m|
\\+& [|0\rangle_{A}\langle0|\otimes\frac{(1-|\gamma_{p}|^{2})^{\frac{3}{2}}}{\sqrt{2}}\sqrt{n+1}[|n\rangle_{B}\langle n+1|+\gamma_{p}|n+1\rangle_{B}\langle n|]\\ \otimes&\frac{(1-|\gamma_{p}|^{2})^{\frac{3}{2}}}{\sqrt{2}}\sqrt{m+1}[\gamma_{p}^{*}|m\rangle_{C}\langle m+1|+|m+1\rangle_{C}\langle m|]+h.c.]
+[|0\rangle_{A}\langle1||n\rangle_{B}\langle n|\\ \otimes&\frac{(1-|\gamma_{p}|^{2})^{\frac{3}{2}}}{\sqrt{2}}\sqrt{m+1}[\gamma_{p}^{*}|m\rangle_{C}\langle m+1|+|m+1\rangle_{C}\langle m|]+h.c.]\\
+&[|1\rangle_{A}\langle0| \otimes\frac{(1-|\gamma_{p}|^{2})^{\frac{3}{2}}}{\sqrt{2}}\sqrt{n+1}[|n\rangle_{B}\langle n+1|+\gamma_{p}|n+1\rangle_{B}\langle n|]\otimes|m\rangle_{C}\langle m|+h.c.]\\
+&(1-|\gamma_{p}|^{2})^{2}{[|1\rangle_{A}\langle1||n\rangle_{B}\langle n||m\rangle_{C}\langle m|]}.
\end{aligned}
\end{equation}
Employing Eq.(\ref{w19}), the coherence of the W state becomes
\begin{equation}\label{w37}
\begin{aligned}
C(\rho_{ABC})=&\frac{1}{3}\big\{2(1-|\gamma_{p}|^{2})^{2}\sum_{n=0}^{\infty}|\gamma_{p}|^{2n+1}\sqrt{n+2}\sqrt{n+1}\\
+&2\sqrt{2}(1-|\gamma_{p}|^{2})^{\frac{3}{2}}\sum_{m=0}^{\infty}|\gamma_{p}|^{2m}\sqrt{m+1}(|\gamma_{p}|+1)\\
+&(1-|\gamma_{p}|^{2})^{3}[\sum_{m=0}^{\infty}|\gamma_{p}|^{2m}\sqrt{m+1}(|\gamma_{p}|+1)]^{2}\big\}.
\end{aligned}
\end{equation}
From Fig.\ref{Fig4}, we can see that quantum coherence of the W state decreases monotonically with the increase of curvature parameter $p$, which means that quantum coherence increases monotonically with the increase of the curvature.  From Fig.\ref{Fig4}, we also see that quantum coherence of the W state  is most severely affected by the curvature effect of de Sitter space for $\nu\rightarrow1/2$ (conformally coupled massless) and $\nu\rightarrow3/2$ ( minimally coupled massless limit). In addition, quantum coherence of the W state is larger than that of the GHZ state, which means that quantum coherence of the W state is more suitable for processing quantum information tasks in the multiverse.

\begin{figure}[htbp]
\centering
\includegraphics[height=1.8in,width=2.0in]{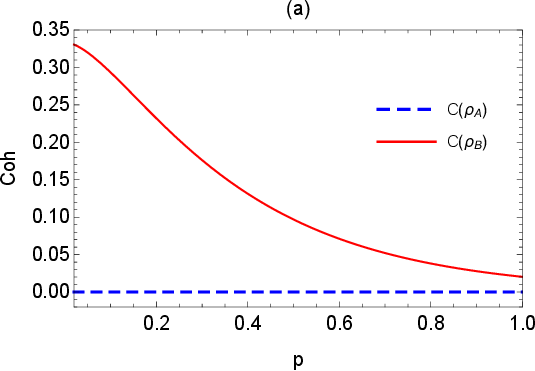}
\includegraphics[height=1.8in,width=2.0in]{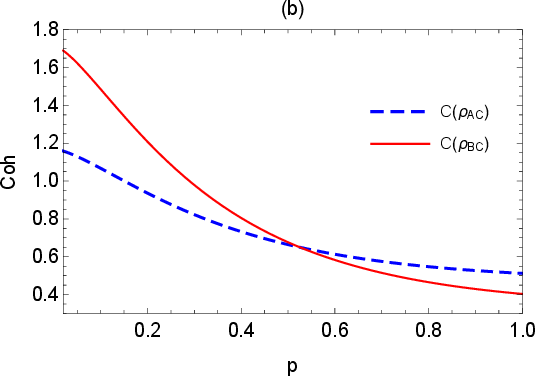}
\includegraphics[height=1.8in,width=2.0in]{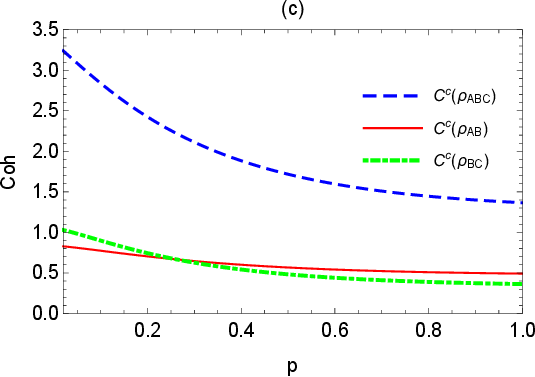}
\caption{ Single particle coherence $C(\rho_{A})$  and $C(\rho_{B})$, bipartite coherence $C(\rho_{BC})$ and $C(\rho_{AB})$, and  correlated coherence $C^{c}(\rho_{AB})$, $C^{c}(\rho_{BC})$ and  $C^{c}(\rho_{ABC})$ of the W state as a function of the curvature parameter $p$ for a fixed $\nu=3/2$.}\label{Fig5}
\end{figure}

We get the density matrix $\rho_{BC}$, $\rho_{AC}$ and $\rho_{AB}$ after tracing over the modes $A$, $B$, or $C$ from the state $\rho_{ABC}$, respectively,

\begin{equation}\label{w38}
\begin{aligned}
\rho_{BC}=&\frac{1}{3}[[(1-|\gamma_{p}|^{2})\sum_{n=0}^{\infty}\gamma_{p}^{2n}|n\rangle_{B}\langle n|\otimes\frac{(1-|\gamma_{p}|^{2})^{2}}{2}\sum_{m=0}^{\infty}\gamma_{p}^{2m}[(m+1)(|m\rangle_{C}\langle m|\\+&|m+1\rangle_{C}\langle m+1|)+\sqrt{m+1}\sqrt{m+2}(\gamma_{p}|m+2\rangle_{C} \langle m|+\gamma_{p}^{*}|m\rangle_{C}\langle m+2|)]+h.c.]\\+&[\frac{(1-|\gamma_{p}|^{2})^{\frac{3}{2}}}{\sqrt{2}}\sum_{n=0}^{\infty}\gamma_{p}^{2n}\sqrt{n+1}
(|n\rangle_{B}\langle n+1|+\gamma_{p}|n+1\rangle_{B}\langle n|) \\ \otimes & \frac{(1-|\gamma_{p}|^{2})^{\frac{3}{2}}}{\sqrt{2}}\sum_{m=0}^{\infty}\gamma_{p}^{2m}\sqrt{m+1}
(|m+1\rangle_{C}\langle m|+\gamma_{p}^{*}|m\rangle_{C}\langle m+1|)+ h.c.]\\+&(1-|\gamma_{p}|^{2})\sum_{n=0}^{\infty}\gamma_{p}^{2n}|n\rangle_{B}\langle n|\otimes(1-|\gamma_{p}|^{2})\sum_{m=0}^{\infty}\gamma_{p}^{2m}|m\rangle_{C}\langle m|],
\end{aligned}
\end{equation}
\begin{equation}\label{w39}
\begin{aligned}
\rho_{AC}=&\frac{1}{3}[|0\rangle_{A}\langle0|\otimes\frac{(1-|\gamma_{p}|^{2})^{2}}{2}\sum_{m=0}^{\infty}\gamma_{p}^{2m}[(m+1)(|m\rangle_{B}\langle m|+|m+1\rangle_{B}\langle m+1|)\\+&\sqrt{m+1}\sqrt{m+2}(\gamma_{p}|m+2\rangle_{B} \langle m|+\gamma_{p}^{*}|m\rangle_{B}\langle m+2|)]\\+&|0\rangle_{A}\langle1| \otimes\frac{(1-|\gamma_{p}|^{2})^{\frac{3}{2}}}{\sqrt{2}}\sum_{m=0}^{\infty}\gamma_{p}^{2m}\sqrt{m+1}
(|m+1\rangle_{B}\langle m|+\gamma_{p}^{*}|m\rangle_{B}\langle m+1|)\\+&|1\rangle_{A}\langle0|\otimes\frac{(1-|\gamma_{p}|^{2})^{\frac{3}{2}}}{\sqrt{2}}\sum_{m=0}^{\infty}\gamma_{p}^{2m}\sqrt{m+1}
(|m+1\rangle_{B}\langle m|+\gamma_{p}^{*}|m\rangle_{B}\langle m+1|)\\+&|0\rangle_{A}\langle0|\otimes
(1-|\gamma_{p}|^{2})\sum_{m=0}^{\infty}\gamma_{p}^{2m}|m\rangle_{B}\langle m|+|1\rangle_{A}\langle1|\otimes(1-|\gamma_{p}|^{2})\sum_{m=0}^{\infty}\gamma_{p}^{2m}|m\rangle_{B}\langle m|],
\end{aligned}
\end{equation}
\begin{equation}\label{w40}
\begin{aligned}
\rho_{AB}=&\frac{1}{3}[|0\rangle_{A}\langle0|\otimes\frac{(1-|\gamma_{p}|^{2})^{2}}{2}\sum_{m=0}^{\infty}\gamma_{p}^{2m}[(m+1)(|m\rangle_{C}\langle m|+|m+1\rangle_{C}\langle m+1|)\\+&\sqrt{m+1}\sqrt{m+2}(\gamma_{p}|m+2\rangle_{C} \langle m|+\gamma_{p}^{*}|m\rangle_{C}\langle m+2|)]\\+&|0\rangle_{A}\langle1| \otimes\frac{(1-|\gamma_{p}|^{2})^{\frac{3}{2}}}{\sqrt{2}}\sum_{m=0}^{\infty}\gamma_{p}^{2m}\sqrt{m+1}
(|m+1\rangle_{C}\langle m|+\gamma_{p}^{*}|m\rangle_{C}\langle m+1|)\\+&|1\rangle_{A}\langle0|\otimes\frac{(1-|\gamma_{p}|^{2})^{\frac{3}{2}}}{\sqrt{2}}\sum_{m=0}^{\infty}\gamma_{p}^{2m}\sqrt{m+1}
(|m+1\rangle_{C}\langle m|+\gamma_{p}^{*}|m\rangle_{C}\langle m+1|)\\+&|0\rangle_{A}\langle0|\otimes
(1-|\gamma_{p}|^{2})\sum_{m=0}^{\infty}\gamma_{p}^{2m}|m\rangle_{C}\langle m|+|1\rangle_{A}\langle1|\otimes(1-|\gamma_{p}|^{2})\sum_{m=0}^{\infty}\gamma_{p}^{2m}|m\rangle_{C}\langle m|].
\end{aligned}
\end{equation}
Using Eq.(\ref{w19}), we obtain the bipartite coherence as
\begin{equation}\label{w41}
\begin{aligned}
C(\rho_{BC})=&\frac{1}{3}\big\{2(1-|\gamma_{p}|^{2})^{2}\sum_{n=0}^{\infty}\sqrt{n+1}\sqrt{n+2}|\gamma_{p}|^{2n+1}\\+&
(1-|\gamma_{p}|^{2})^{3}\big[\sum_{n=0}^{\infty}|\gamma_{p}|^{2n}\sqrt{n+1}(|\gamma_{p}|+1)\big]^{2}\big\},
\end{aligned}
\end{equation}
\begin{equation}\label{w42}
\begin{aligned}
C(\rho_{AC})=C(\rho_{AB})=&\frac{1}{3}[(1-|\gamma_{p}|^{2})^{2}\sum_{n=0}^{\infty}\sqrt{n+1}\sqrt{n+2}|\gamma_{p}|^{2n+1}\\+&
\sqrt{2}(1-|\gamma_{p}|^{2})^{\frac{3}{2}}\sum_{n=0}^{\infty}|\gamma_{p}|^{2n}\sqrt{n+1}(1+|\gamma_{p}|)].
\end{aligned}
\end{equation}
Similarly, we get the density matrices $\rho_{A}, \rho_{B}$, and $\rho_{C}$ of  single particle system as
\begin{equation}\label{w43}
\begin{aligned}
\rho_{A}=\frac{1}{3}(2|0\rangle_{A}\langle0|+|1\rangle_{A}\langle1|),
\end{aligned}
\end{equation}
\begin{equation}\label{w44}
\begin{aligned}
\rho_{B}=&\frac{1}{3}[2(1-|\gamma_{p}|^{2})\sum_{m=0}^{\infty}\gamma_{p}^{2m}|m\rangle_{B}\langle m|+\frac{(1-|\gamma_{p}|^{2})^{2}}{2}\sum_{m=0}^{\infty}\gamma_{p}^{2m}[(m+1)(|m\rangle_{B}\langle m|\\+&|m+1\rangle_{B}\langle m+1|)+\sqrt{m+1}\sqrt{m+2}(\gamma_{p}|m+2\rangle_{B} \langle m|+\gamma_{p}^{*}|m\rangle_{B}\langle m+2|)]],
\end{aligned}
\end{equation}
\begin{equation}\label{w45}
\begin{aligned}
\rho_{C}=&\frac{1}{3}[2(1-|\gamma_{p}|^{2})\sum_{m=0}^{\infty}\gamma_{p}^{2m}|m\rangle_{C}\langle m|+\frac{(1-|\gamma_{p}|^{2})^{2}}{2}\sum_{m=0}^{\infty}\gamma_{p}^{2m}[(m+1)(|m\rangle_{C}\langle m|\\+&|m+1\rangle_{C}\langle m+1|)+\sqrt{m+1}\sqrt{m+2}(\gamma_{p}|m+2\rangle_{C} \langle m|+\gamma_{p}^{*}|m\rangle_{C}\langle m+2|)]].
\end{aligned}
\end{equation}
The corresponding single particle coherence reads
\begin{equation}\label{w46}
\begin{aligned}
C(\rho_{A})=0,
\end{aligned}
\end{equation}
\begin{equation}\label{w47}
\begin{aligned}
C(\rho_{B})=C(\rho_{C})=&\frac{(1-|\gamma_{p}|^{2})^{2}}{3}\sum_{n=0}^{\infty}\sqrt{n+1}\sqrt{n+2}|\gamma_{p}|^{2n+1}.
\end{aligned}
\end{equation}
According to Eq.(\ref{w28}), we can obtain the  correlated coherence as
\begin{equation}\label{w48}
\begin{aligned}
C^{c}(\rho_{ABC})=&\frac{1}{3}\big\{2\sqrt{2}(1-|\gamma_{p}|^{2})^{\frac{3}{2}}\sum_{m=0}^{\infty}|\gamma_{p}|^{2m}\sqrt{m+1}(|\gamma_{p}|+1)\\
+&(1-|\gamma_{p}|^{2})^{3}[\sum_{m=0}^{\infty}|\gamma_{p}|^{2m}\sqrt{m+1}(|\gamma_{p}|+1)]^{2}\big\},
\end{aligned}
\end{equation}
\begin{equation}\label{w49}
\begin{aligned}
C^{c}(\rho_{AB})=C^{c}(\rho_{AC})=&\frac{\sqrt{2}(1-|\gamma_{p}|^{2})^{\frac{3}{2}}}{3}\sum_{n=0}^{\infty}|\gamma_{p}|^{2n}\sqrt{n+1}(|\gamma_{p}|+1),
\end{aligned}
\end{equation}
\begin{equation}\label{w50}
\begin{aligned}
C^{c}(\rho_{BC})=&\frac{(1-|\gamma_{p}|^{2})^{3}}{3}\big[\sum_{n=0}^{\infty}|\gamma_{p}|^{2n}\sqrt{n+1}(|\gamma_{p}|+1)\big]^{2}.
\end{aligned}
\end{equation}
From Fig.\ref{Fig5}, we can see that the properties of the single particle coherence $C(\rho_{A})$  and $C(\rho_{B})$, bipartite coherence $C(\rho_{BC})$ and $C(\rho_{AB})$, and  correlated coherence $C^{c}(\rho_{AB})$, $C^{c}(\rho_{BC})$ and  $C^{c}(\rho_{ABC})$ in  the W state are similar to those in the GHZ state.

The relationship of quantum coherence for the W state is still unclear in the multiverse. Through direct calculation,  we obtain a distribution relationship between the correlated coherence of the tripartite W state as
\begin{equation}\label{w51}
\begin{aligned}
C^{c}(\rho_{ABC})=&C^{c}(\rho_{AB})+C^{c}(\rho_{BC})+C^{c}(\rho_{AC}).
\end{aligned}
\end{equation}
From Eq.(\ref{w51}), we can see that the correlated coherence of the W state exists essentially in the form of bipartite correlated coherence, showing that the correlated coherence of the tripartite system is equal to the sum of all the  bipartite  correlated coherence in the multiverse. In addition, we find that Eqs.(\ref{w12}) and (\ref{w13}) in this paper are very similar to Eqs.(14) and (16) for $q_R=1/\sqrt{2}$ in reference \cite{L520}. Therefore, the properties of multipartite coherence in de Sitter space  are similar to those beyond the single-mode approximation in Rindler spacetime.

\section{N-partite coherence in multiverse  \label{GSCDGE}}
In this section, we will discuss the extension of the tripartite systems to the N-partite systems ($N$ $ \geq 3$).  The N-partite GHZ state and N-partite W state can be
written as
\begin{eqnarray}\label{w52}
|GHZ\rangle_{123\cdot\cdot\cdot N}&=&\frac{1}{\sqrt{2}}(|0\rangle_{\rm BD, 1}|0\rangle_{\rm BD, 2}...|0\rangle_{\rm BD, (N-1)}|0\rangle_{\rm BD, N} \nonumber\\
&+&|1\rangle_{\rm BD, 1}|1\rangle_{\rm BD, 2}...|1\rangle_{\rm BD, (N-1)}|1\rangle_{\rm BD, N}),
\end{eqnarray}
\begin{equation}\label{w53}
\begin{aligned}
|W\rangle_{123\cdot\cdot\cdot N}=&\frac{1}{N}(|1\rangle_{\rm BD, 1}|0\rangle_{\rm BD, 2}..|0\rangle_{\rm BD, (N-1)}|0\rangle_{\rm BD, N}+|0_{\rm BD, 1}|1\rangle_{\rm BD, 2}...|0_{\rm BD, (N-1)}|0\rangle_{\rm BD, N}\\&+\cdot\cdot\cdot+|0_{\rm BD,1}|0\rangle_{\rm BD, 2}...|0\rangle_{\rm BD, (N-1)}|1\rangle_{\rm BD, N}),
\end{aligned}
\end{equation}
where the mode $i$ ($i=1, 2, ..., N$)) is observed by observer $O_i$ in  de Sitter space $\rm BD_{i}$. Now we assume that $n$ ($2\leq n\leq N-1$) observers are  in the different $L$ regions of the expanded de Sitter spaces  and $N-n$ observers are   in global charts of the other de Sitter spaces. By a series of calculations,  we obtain N-partite coherence of the GHZ and W states as
\begin{equation}\label{w54}
\begin{aligned}
C(GHZ)=&\frac{1}{2}\bigg\{2\bigg[\frac{(1-|\gamma_{p}|^{2})^{\frac{3}{2}}}{\sqrt{2}}\sum_{m=0}^{\infty}
|\gamma_{p}|^{2m}\sqrt{m+1}(|\gamma_{p}|+1)\bigg]^{n}\\+&\bigg[1+(1-|\gamma_{p}|^{2})^{2}\sum_{m=0}^{\infty}
|\gamma_{p}|^{2m+1}\sqrt{m+1}\sqrt{m+2}\bigg]^{n}-1\bigg\},
\end{aligned}
\end{equation}
\begin{equation}\label{w55}
\begin{aligned}
C(W)=&\frac{1}{N}\bigg\{n(1-|\gamma_{p}|^{2})^{2}\sum_{m=0}^{\infty}\sqrt{m+1}\sqrt{m+2}|\gamma_{p}|^{2m+1}\\+&
2n(N-n)\frac{(1-|\gamma_{p}|^{2})^{\frac{3}{2}}}{\sqrt{2}}\sum_{m=0}^{\infty}
|\gamma_{p}|^{2m}\sqrt{m+1}(|\gamma_{p}|+1)\\+&n(n-1)[\frac{(1-|\gamma_{p}|^{2})^{\frac{3}{2}}}{\sqrt{2}}\sum_{m=0}^{\infty}
|\gamma_{p}|^{2m}\sqrt{m+1}(|\gamma_{p}|+1)]^{2}\\+&(N-n)(N-n-1)\bigg\}.
\end{aligned}
\end{equation}
After tedious but straightforward calculations, the  correlated coherence of the GHZ and W states reads
\begin{equation}\label{w56}
\begin{aligned}
C^{c}(GHZ)=&\frac{1}{2}\bigg\{2\bigg[\frac{(1-|\gamma_{p}|^{2})^{\frac{3}{2}}}{\sqrt{2}}\sum_{m=0}^{\infty}
|\gamma_{p}|^{2m}\sqrt{m+1}(|\gamma_{p}|+1)\bigg]^{n}\\+&\bigg[1+(1-|\gamma_{p}|^{2})^{2}\sum_{m=0}^{\infty}
|\gamma_{p}|^{2m+1}\sqrt{m+1}\sqrt{m+2}\bigg]^{n}-1\\-&n(1-|\gamma_{p}|^{2})^{2}\sum_{m=0}^{\infty}
|\gamma_{p}|^{2m+1}\sqrt{m+1}\sqrt{m+2}\bigg\},
\end{aligned}
\end{equation}
\begin{equation}\label{w57}
\begin{aligned}
C^{c}(W)=&\frac{1}{N}\bigg[2n(N-n)\frac{(1-|\gamma_{p}|^{2})^{\frac{3}{2}}}{\sqrt{2}}\sum_{m=0}^{\infty}
|\gamma_{p}|^{2m}\sqrt{m+1}(|\gamma_{p}|+1)\\+&n(n-1)[\frac{(1-|\gamma_{p}|^{2})^{\frac{3}{2}}}{\sqrt{2}}\sum_{m=0}^{\infty}
|\gamma_{p}|^{2m}\sqrt{m+1}(|\gamma_{p}|+1)]^{2}\\+&(N-n)(N-n-1)\bigg].
\end{aligned}
\end{equation}
From Eqs.(\ref{w54})-(\ref{w57}), we can  see that N-partite (correlated )  coherence of the GHZ state depends only on the $n$ observers who are in the different $L$ regions of the expanded de Sitter spaces, while  N-partite (correlated ) coherence of the W state depends not only on the $n$ observers  but also on the initial  $N$ observers.

\begin{figure}
\begin{minipage}[t]{0.5\linewidth}
\centering
\includegraphics[width=2.6in]{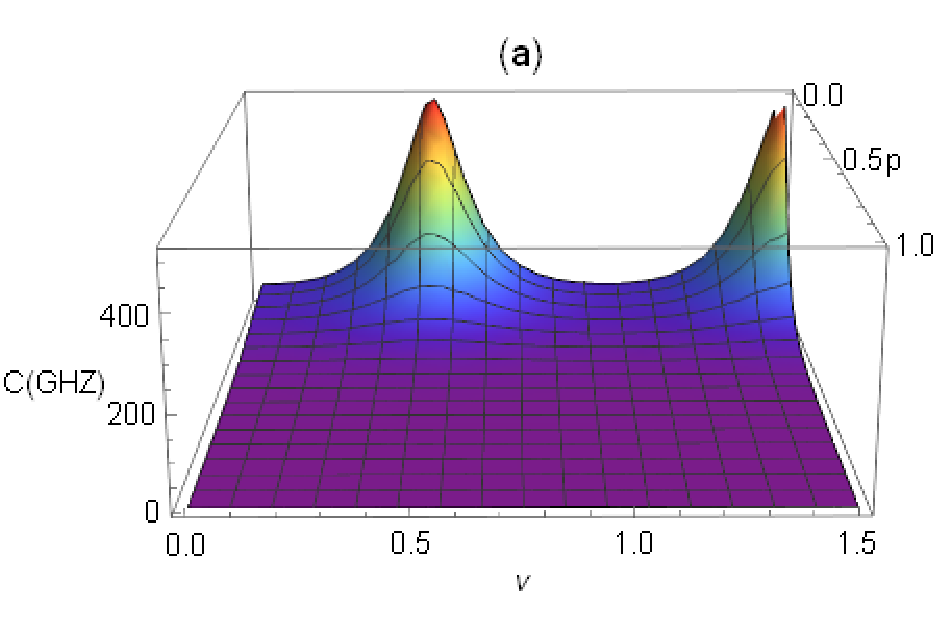}
\label{fig5a}
\end{minipage}%
\begin{minipage}[t]{0.5\linewidth}
\centering
\includegraphics[width=2.6in]{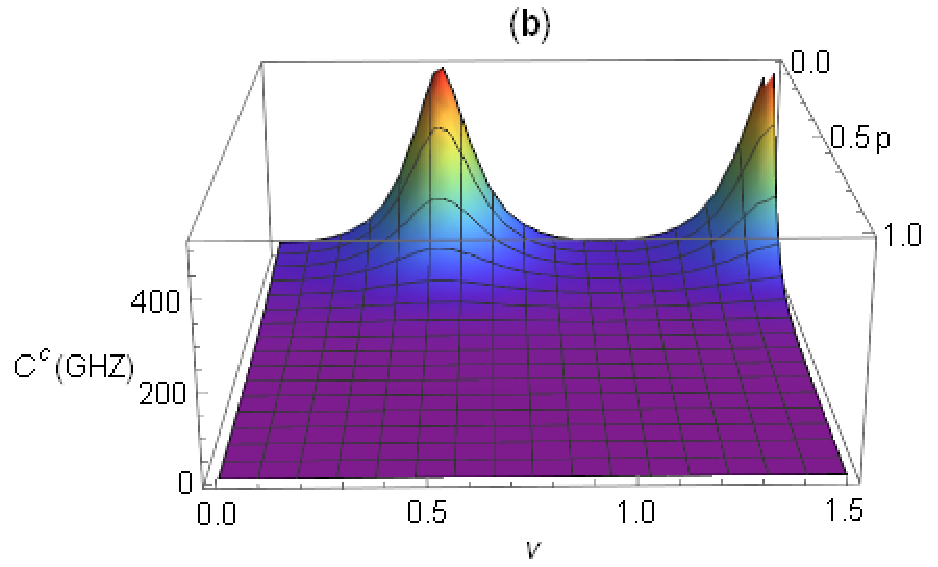}
\label{fig5b}
\end{minipage}%

\begin{minipage}[t]{0.5\linewidth}
\centering
\includegraphics[width=2.6in]{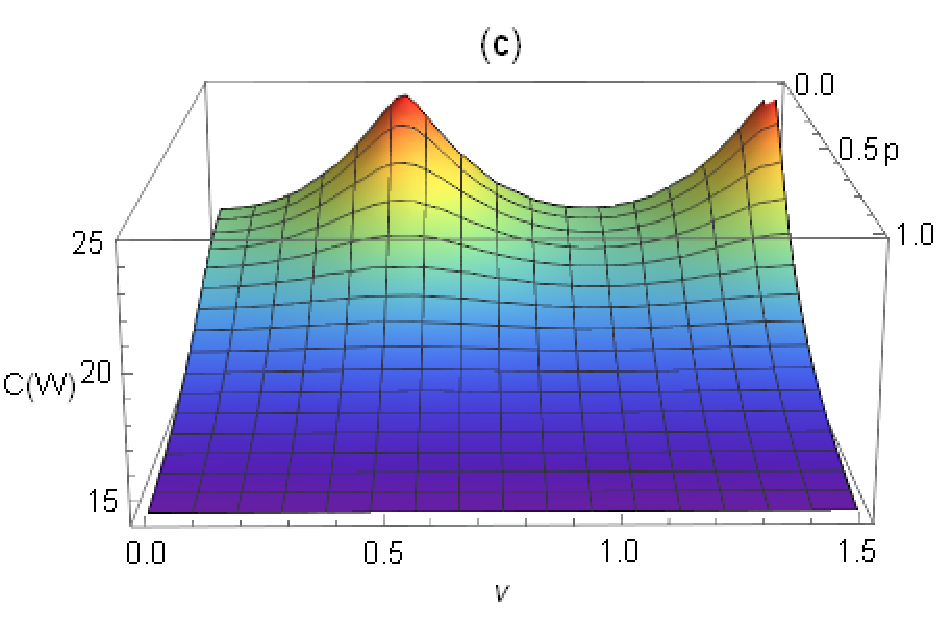}
\label{fig5c}
\end{minipage}%
\begin{minipage}[t]{0.5\linewidth}
\centering
\includegraphics[width=2.6in]{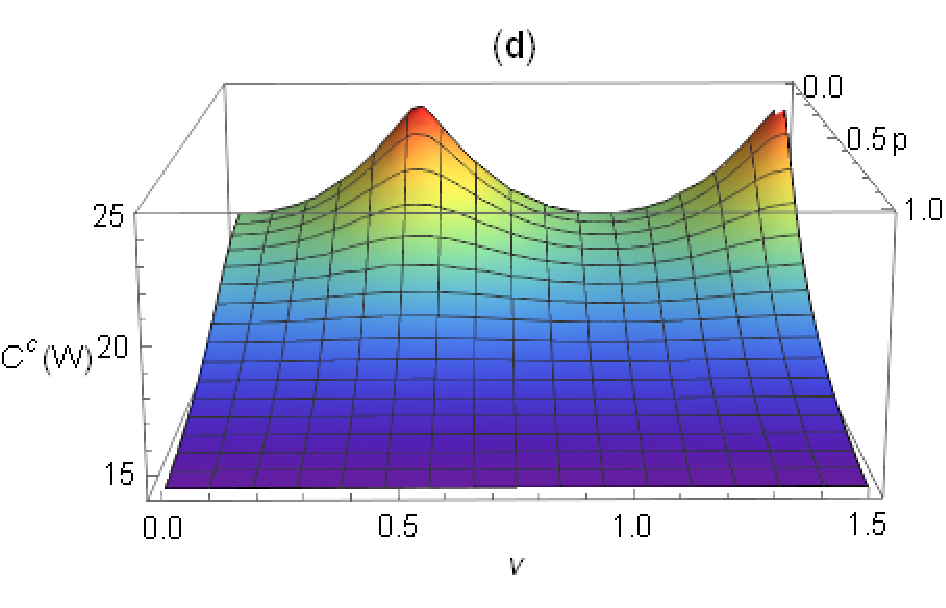}
\label{fig5d}
\end{minipage}%
\caption{N-partite  coherence and   correlated coherence of the GHZ and W states as  functions of the mass parameter $\nu$ and the curvature parameter $p$ for fixed  $n$=10 and $N$=20.}
\label{Fig6}
\end{figure}

\begin{figure}
\begin{minipage}[t]{0.5\linewidth}
\centering
\includegraphics[width=3in]{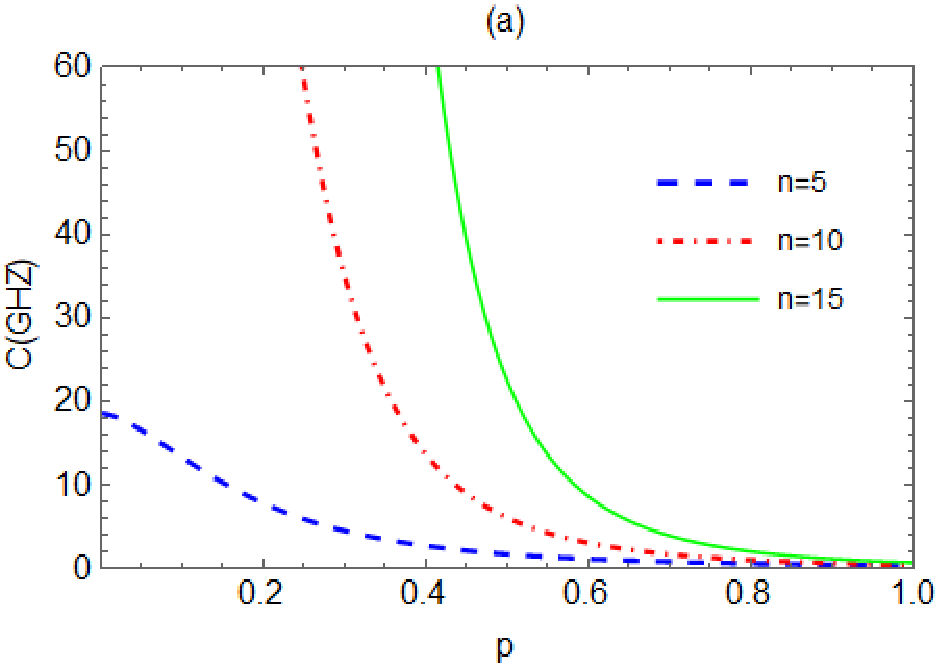}
\label{fig6a}
\end{minipage}%
\begin{minipage}[t]{0.5\linewidth}
\centering
\includegraphics[width=3in]{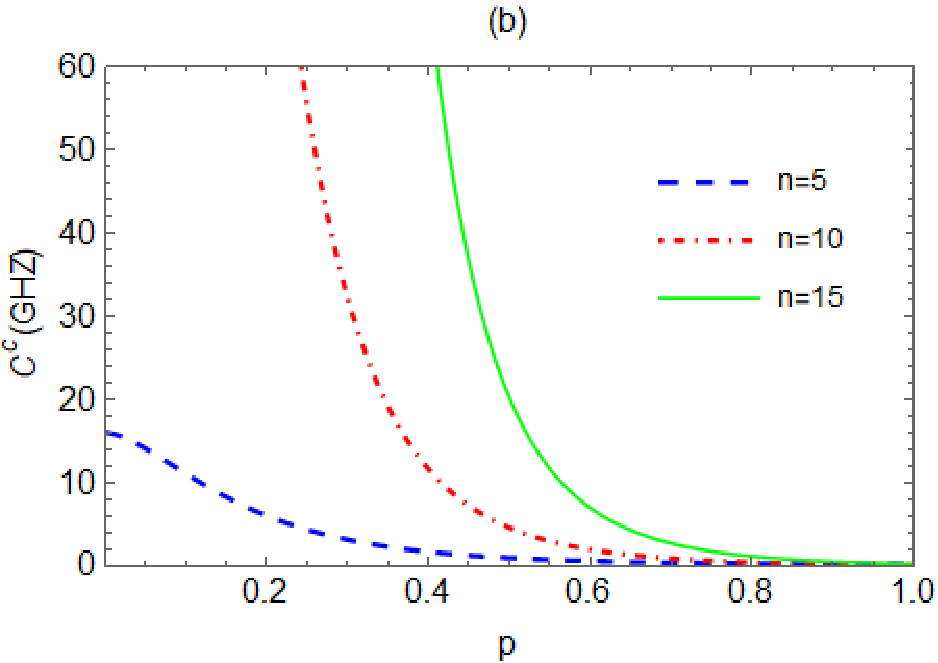}
\label{fig6b}
\end{minipage}%

\begin{minipage}[t]{0.5\linewidth}
\centering
\includegraphics[width=3in]{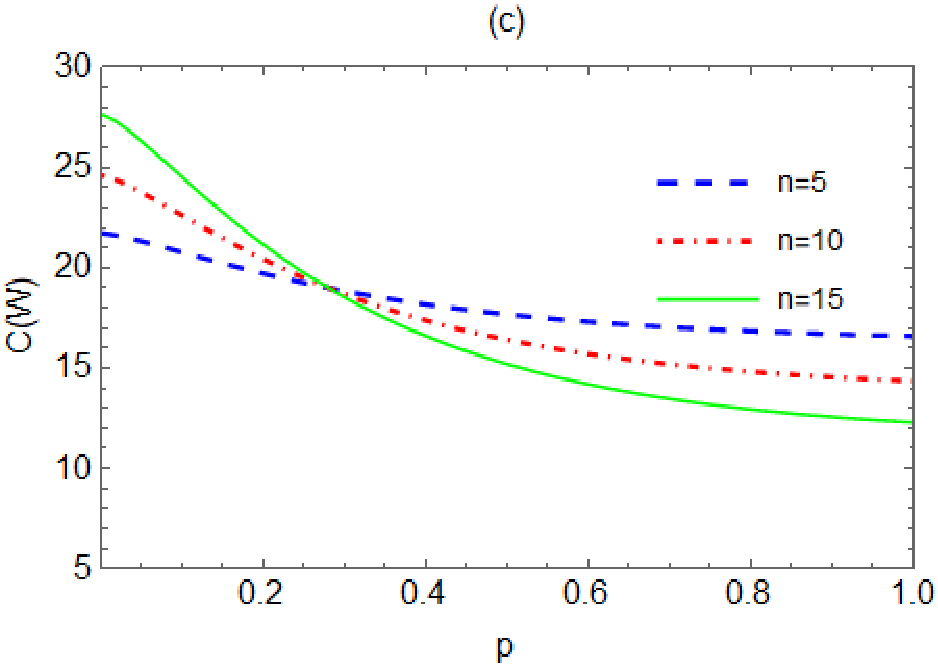}
\label{fig6c}
\end{minipage}%
\begin{minipage}[t]{0.5\linewidth}
\centering
\includegraphics[width=3in]{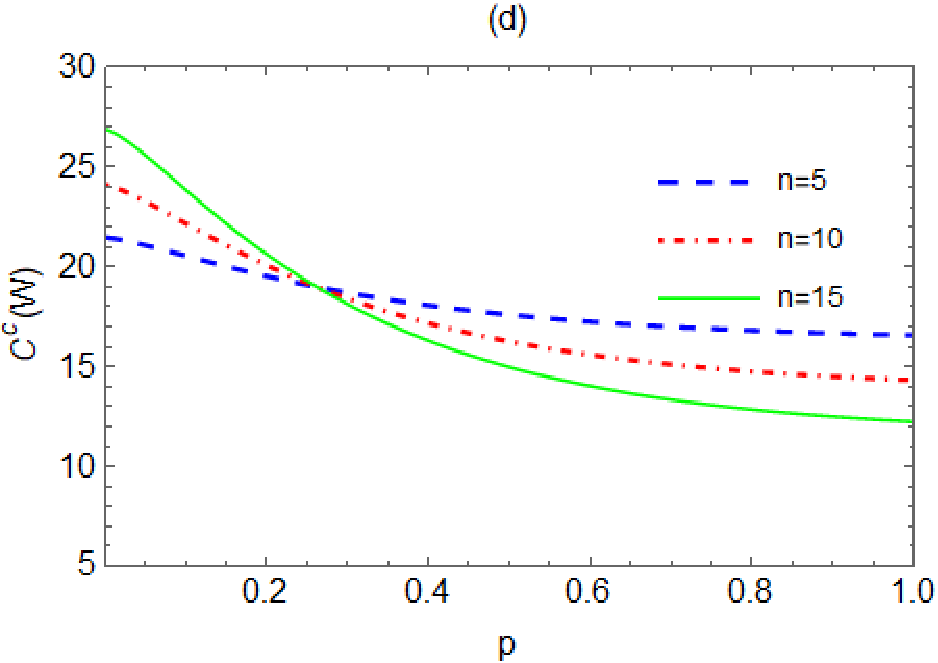}
\label{fig6d}
\end{minipage}%
\caption{N-partite  coherence  and   correlated coherence  of the GHZ and W states as a function of the curvature parameter $p$   for  fixed $\nu= 3/2$, $N$ = 20 and different $n$.}
\label{Fig7}
\end{figure}

\begin{figure}
\begin{minipage}[t]{0.5\linewidth}
\centering
\includegraphics[width=2.6in]{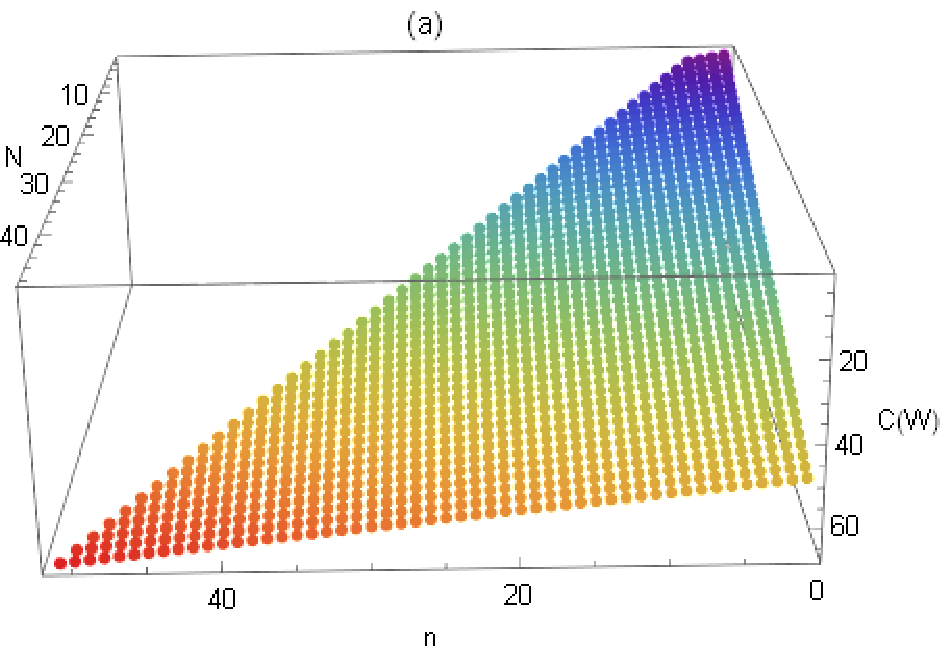}
\label{fig7a}
\end{minipage}%
\begin{minipage}[t]{0.5\linewidth}
\centering
\includegraphics[width=2.6in]{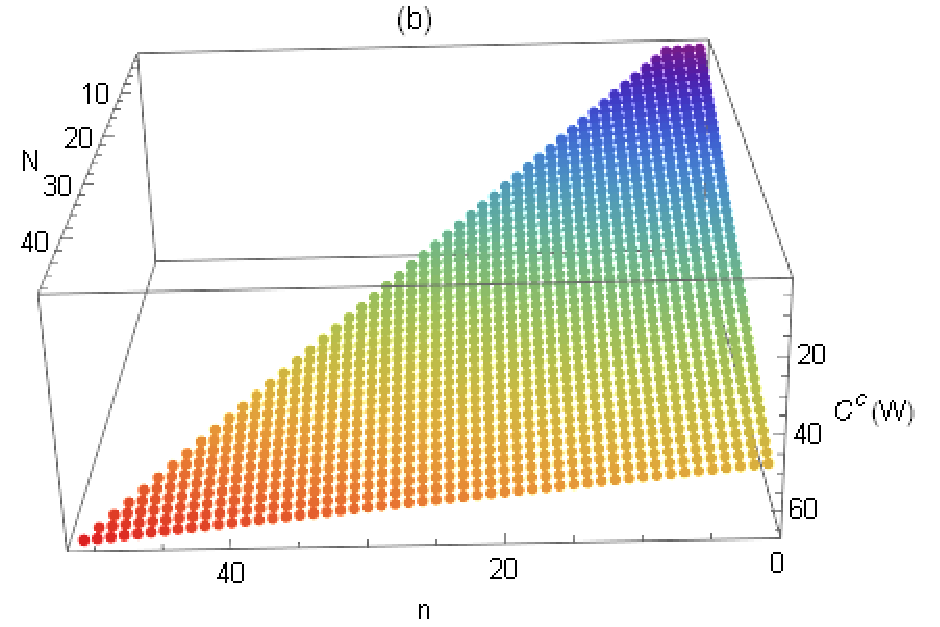}
\label{fig7b}
\end{minipage}%

\begin{minipage}[t]{0.5\linewidth}
\centering
\includegraphics[width=2.6in]{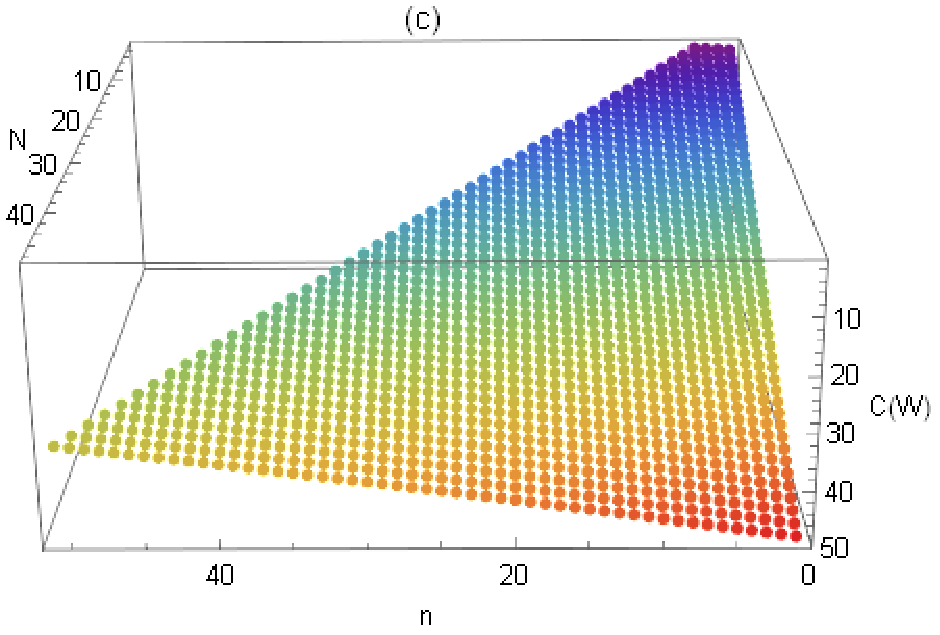}
\label{fig7c}
\end{minipage}%
\begin{minipage}[t]{0.5\linewidth}
\centering
\includegraphics[width=2.6in]{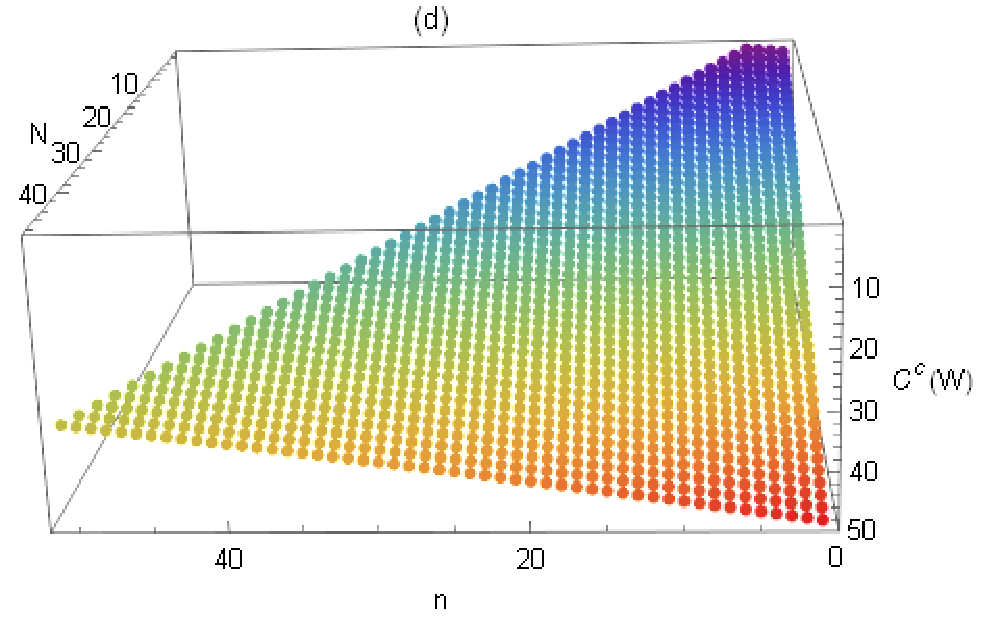}
\label{fig7d}
\end{minipage}%
\caption{N-partite  coherence  and   correlated coherence of the W state as functions of $n$ and $N$ for  $p=0.1$ (a) and (b), $p=0.6$ (c) and (d). The
parameter $\nu$ is fixed as $\nu=3/2$.}
\label{Fig8}
\end{figure}

From Fig.\ref{Fig6}, we find that the curvature effect enhances  N-partite (correlated) coherence. Therefore, the curvature effect is beneficial to N-partite (correlated)  coherence in the multiverse consisting of many  de Sitter spaces. From Fig.\ref{Fig7} (a)-(b), we can see  that N-partite  (correlated) coherence of the GHZ state  increases monotonically with increasing $n$ for any curvature parameter $p$.
From Fig.\ref{Fig7} (c)-(d), we can also see  that, with the increase of $n$,
N-partite  (correlated) coherence of the W state for smaller curvature parameter $p$ increases monotonically, while for larger curvature parameter $p$ decreases monotonically in the multiverse. From Fig.\ref{Fig8},  we can show that   N-partite  (correlated) coherence of the W state increases monotonically with the increase of $N$ for any $p$ and $n$, while N-partite  (correlated) coherence of the GHZ state is independent of $N$.
We can also see that  N-partite  (correlated) coherence of the W state increases or decreases  monotonically with the growth of $n$ depending on the curvature parameter $p$.

Next, we study the distribution relationship of coherence for the N-partite systems in the multiverse. For N-partite  W state, the distribution relationship of the correlated coherence  can be written as
\begin{equation}\label{w58}
\begin{aligned}
\frac{n(n-1)}{2}C^{c}(\rho_{O} \rho_{O} )+n(N-n)C^{c}(\rho_{O}\rho_{X} )+\frac{(N-n)(N-n-1)}{2}C^{c}(\rho_{X}\rho_{X})=C^{c}(W).
\end{aligned}
\end{equation}
Here, $$C^{c}(\rho_{O}\rho_{O} )=\frac{2}{N}[\frac{(1-|\gamma_{p}|^{2})^{\frac{3}{2}}}{\sqrt{2}}\sum_{m=0}^{\infty}
|\gamma_{p}|^{2m}\sqrt{m+1}(|\gamma_{p}|+1)]^{2}$$ denotes the bipartite correlated coherence of the modes corresponding to  two observers  in the $L$ regions of the expanded de Sitter spaces; $$C^{c}(\rho_{O}\rho_{X})= \frac{2}{N}\frac{(1-|\gamma_{p}|^{2})^{\frac{3}{2}}}{\sqrt{2}}\sum_{m=0}^{\infty}
|\gamma_{p}|^{2m}\sqrt{m+1}(|\gamma_{p}|+1)$$ denotes the bipartite correlated coherence of the modes corresponding to  two observers   who are,  respectively, in the $L$ region of the expanded de Sitter space and  the global chart of  de Sitter space; $C^{c}(\rho_{X}\rho_{X})= \frac{2}{N}$ denotes the bipartite correlated coherence of the modes corresponding to  two observers  in the global chart of  de Sitter spaces. From  Eq.(\ref{w58}), we can see that the total correlated coherence of N-partite W state is equal to the sum of all bipartite correlated coherence in the multiverse.

\section{ Conclutions  \label{GSCDGE}}
The  curvature effect on quantum coherence of  tripartite GHZ and W states in the multiverse  has been  investigated.  It is
shown that tripartite coherence increases with the increase of the curvature, meaning that the curvature effect is beneficial to quantum coherence in the multiverse.
However, with the growth of the curvature, quantum entanglement gradually vanishes and quantum discord gradually decreases to a fixed value at the limit of infinite curvature in de Sitter space  \cite{L25,L26,L27,L28}.  We find that quantum coherence of the GHZ and W states is sensitive to the curvature effect in the limit of
conformal and massless scalar fields. Interestingly, the curvature effect can generate bipartite quantum coherence and single-partite quantum coherence.
We obtain a distribution relationship of correlated coherence for the W state in the multiverse. In other words, tripartite  correlated coherence of the W state is equal to the sum of all bipartite correlated coherence.

We have extended the investigation from the tripartite systems to N-partite systems in the multiverse. Unlike tripartite coherence, N-partite coherence of the W state depends not only on the curvature and mass parameters, but also on the $n$ particles in the $L$ regions of the de Sitter spaces and  the initial  $N$ particles.  However, N-partite coherence of the GHZ state is independent of the $N$.
We find that with the increase of the $n$, N-partite coherence of the GHZ state increases monotonically for any  curvature, while N-partite coherence of the W state increases or decreases monotonically, depending on the curvature. We extend the tripartite  distribution relationship to the N-partite distribution relationship. This indicates that the total correlated coherence of N-partite W state is essentially bipartite types.
Based on the above arguments, we expect that multipartite coherence will be able to provide some evidence for the existence of the multiverse.

\begin{acknowledgments}
This work is supported by the National Natural
Science Foundation of China (Grant Nos. 12205133 and 1217050862), LJKQZ20222315 and JYTMS20231051.
\end{acknowledgments}


\end{document}